\documentclass[twocolumn,showpacs,showkeys,amsmath,amssymb,floatfix,aps,prc]{revtex4}
\newcommand{\bm}{\bibitem}
\usepackage{dcolumn}
\usepackage{supertabular}
\usepackage{color,graphicx}
\usepackage[bookmarksopen]{hyperref}

\def  \a    {\alpha}
\def  \b    {\beta }
\def  \g    {\gamma}
\def  \G    {\Gamma}
\def  \l    {\lambda}
\def  \L    {\Lambda}
\def  \o    {\omega}
\def  \O    {\Omega}
\def  \s    {\sigma}
\def  \S    {\Sigma}
\def  \p    {\pi}

\def  \m    {\mu}
\def  \n    {\nu}

\def  \t    {\tau}
\def  \v    {\vec}
\def  \f    {\frac}
\def  \lt   {\left}
\def  \rt   {\right}
\def  \th   {\theta}

\def  \ra   {\rightarrow}

\def  \dt   {\delta}
\def  \Dt   {\Delta}
\def  \ep   {\epsilon}
\def  \z    {\zeta}

\def  \lv   {\left\vert}
\def  \rv   {\right\vert}
\def  \be   {\begin{equation}}
\def  \ee   {\end{equation}}
\def  \ba   {\begin{array}}
\def  \ea   {\end{array}}
\def  \bea  {\begin{eqnarray}}
\def  \eea  {\end{eqnarray}}
\def  \nn   {\nonumber}
\def  \bd   {\begin{displaymath}}
\def  \ed   {\end{displaymath}}
\def  \bse  {\begin{subequations}}
\def  \ese  {\end{subequations}}
\def  \bwt  {\begin{widetext}}
\def  \ewt  {\end{widetext}}

\begin{document}

\title{Effects of Dirac sea on pion propagation in asymmetric nuclear matter}

\author{Subhrajyoti Biswas and Abhee K. Dutt-Mazumder}
\address{Saha Institute of Nuclear Physics,
1/AF Bidhannagar, Kolkata-700 064, INDIA}

\medskip

\begin{abstract}
 We study pion propagation in asymmetric nuclear matter (ANM). One of the 
 interesting consequences of pion propagation in ANM is the mode splitting 
 for the different charged states of pions. First we describe the
 pion-nucleon dynamics using 
 the non-chiral model where one starts with pseudoscalar (PS) $\pi$N coupling 
 and the pseudovector (PV) representation is obtained via suitable 
 non-linear field transformations. For both of these cases the effect of the 
 Dirac sea is estimated. Subsequently, we present results using the 
 chiral effective Lagrangian where the short-distance behavior (Dirac
 vacuum) is included by re-defining the field parameters as done in the 
 modern effective field theory approach developed recently. The results
 are compared with the previous calculations for the case of symmetric
 nuclear matter (SNM). Closed form analytical results are presented for
 the effective pion masses and dispersion relations by making hard
 nucleon loop (HNL) approximation and suitable density expansion.
\end{abstract}

\vspace{0.08 cm}

\date{\today}



\maketitle

\section{introduction}

 Pions in nuclear physics assume a special status. It is responsible for the
 spin-isospin dependent long range part of the nuclear force. In
 addition, there are variety of physical phenomena related to the pion 
 propagation in nuclear  matter. One of the fascinating ideas in
 relation to the pion-nucleon dynamics in nuclear matter is the pion 
 condensation \cite{Migdal78}. This might happen
 if there exists space like zero energy excitation of pionic modes. The
 short-range correlation, on the other hand, removes such a
 possibility at least at densities near the saturation densities. In
 the context of relativistic heavy ion collisions (RHIC), the
 importance of medium modified pion spectrum
 was discussed by Mishustin, where it was shown that due to the lowering of
 energy, pion, in nuclear matter, might carry a bulk amount of entropy
 \cite{Mishustin80}. Subsequently, Gyulassy and Greiner studied pionic
 instability in great detail in the context of RHIC \cite{Gyulassy77}.
 The production of pionic modes in nuclear collisions was also
 discussed in \cite{Brown89}.\\

 In experiments medium dependent pion dispersion relation can also be
 probed via the measurements of dilepton invariant mass spectrum. The
 lepton pairs produced with invariant mass near the $\rho$ pole are
 sensitive to the slope of the pion
 dispersion relation in matter \cite{Xia88}. Particularly the softening of
 momentum dependence of the pion dispersion relation in matter leads to higher
 yield of dileptons. Gale and Kapusta were first to realize that the
 in-medium pion dynamics can be studied by measuring lepton pair productions
 \cite{Gale87}. Most of the earlier studies of in-medium pion properties were
 performed in the non-relativistic frame work 
 \cite{Oset82,Migdal90,Dmitriev85}. A quasi-relativistic approach was
 taken in \cite{Henning94,Korpa95,Helgesson95}
 where the calculations were extended to finite temperature. In particular,
 \cite{Helgesson95} discusses various non-collective modes with the possibility
 of pion condensation. In \cite{Xia88}, on the other hand, the dilepton
 production rates were calculated using non-relativistic pion dispersion
 relations. Ref.\cite{Liu97} treated the problem relativistically but
 free Fermi gas model was used, while in \cite{Herbert92} pion
 propagation was studied by extending the Walecka model
 \cite{Walecka74} including delta baryon. In recent years, there has
 been significant progress to calculate dilepton production rates involving
 pionic properties in a more realistic framework
 \cite{Helgesson95,Xia88,Rapp00,Gale87,Wong94,Chanfray93}.\\

 In the present paper we study pion dispersion relations in ANM using
 relativistic models. This is important as most of the calculations, as 
 mentioned above, are either restricted to SNM or performed in
 the non-relativistic framework. Here we focus on the propagating modes of
 various charged states of pions which are non degenerate in ANM. The 
 importance of relativistic corrections and density dependent pion
 mass splitting in ANM in the context of deriving pion-nucleus optical 
 potential was discussed in \cite{Weise01}. The formalism adopted in 
 \cite{Weise01} was that of chiral perturbation theory. Recently, in
 the context of astrophysics, pionic
 properties in ANM has also been studied by involving Nambu-Jona-Lasinio model
 \cite{Costa03,Costa04}. Motivated by \cite{Weise01} present authors revisited
 the problem in ref.\cite{Biswas06} where not only the static self-energy
 responsible for the mass splitting but the full dispersion relations for the
 various charged states of pions were calculated after performing relevant
 density expansion in terms of the Fermi momentum. However, in our
 previous work \cite{Biswas06}, pions were included via straight
 forward PV coupling in the Walecka model \cite{Walecka74} which
 renders the theory non-renormalizable.
 Although the problem of non-renormalizibility could be avoided by
 considering the PS $\p$N coupling. This, on the other hand,
 fails to account for the pion-nucleon phenomenology.\\

 Historically, the extension of the Walecka model  to include
 the isovector $\pi$ and $\rho$ meson for the realistic description
 of dense nuclear matter (DNM) while retaining the renormalizibility
 of the theory was first made by Serot \cite{Serot79}. However, in
 this work, the calculation was restricted only  to the mean field
 level which gives to rise tachyonic mode for pions
 even at density as low as $0.1\rho_0$, where $\rho_0$ denotes normal
 nuclear matter (NNM) density \cite{Kapusta81}. Such a non-propagating
 mode for the pions can be removed by extending the calculation beyond
 the mean field level as showed by Kapusta \cite{Kapusta81}. This, in
 effect, means inclusion of the $\pi$-$NN$ loop while calculating the 
 in-medium dressed propagator for the pion. This model has an added
 advantage because of the presence of $\p$-$\s$ coupling in addition
 to the usual PS coupling of the pion with the nucleons which is
 responsible for the generation of small s-wave pion
 nucleon interaction in vacuum. This is consistent with the observed 
 characteristics of the pion-nucleon interaction which is dominated by
 the p-wave scattering while the $s$-wave scattering length is almost
 zero. In matter, however, as argued in \cite{Kapusta81,Matsui82},
 such subtle cancellation does not occur resulting in a unrealistically 
 large mass for the pions in matter. To circumvent this problem it was 
 suggested in \cite{Kapusta81} to use the pseudovector coupling even
 though it makes the theory non-renormalizable.\\

 The theoretical challenge, therefore, is to construct a model with $\p$N PV
 interaction which preserves the renormalizibility of the theory. This was
 accomplished in ref.~\cite{Matsui82} following the technique
 developed by Weinberg \cite{Weinberg67,Weinberg68,Weinberg79}
 and Schwinger \cite{Schwinger67}. 
 Here one starts with the PS coupling and subsequently
 invokes non-linear field transformations to obtain PV representation. 
 Unlike straight forward inclusion of PV interaction in this
 method one requires only finite number of counter terms which makes the theory
 renormalizable. We, here, start with this model developed by  Matsui and 
 Serot \cite{Matsui82} to study the pion propagation in ANM. 
 Clearly, the model adopted here is different from what we had invoked in our
 previous work \cite{Biswas06}. Furthermore, in \cite{Biswas06}, for
 the determination of pion self-energy in matter
 only the scattering from the Fermi sphere was considered and the vacuum part
 was completely ignored. The latter gives rise to a large contribution to the
 pion self-energy in presence of strong scalar density ($\rho_s$).\\ 

 The above mentioned model has various shortcomings too. In fact, the
 ref.\cite{Matsui82} itself discusses its limitations in 
 describing many body $\pi N$ dynamics. For example, the successful description
 of the saturation properties of nuclear matter in this scheme requires higher
 scalar mass which gives rise to larger in-medium nucleon mass compared to the
 MFT. In addition, it also fails to account for the observed pion-nucleus
 scattering length at finite density \cite{Matsui82}. In the same work, chiral
 $\pi$-$\sigma$ model has also been discussed to  which we shall come later.
 In the end, we present  results calculated using this non-chiral
 model together with what we obtain from a chirally invariant Lagrangian.\\

 In \cite{Biswas06} we have discussed another
 interesting possibility of the density driven $\pi$-$\eta$ mixing in
 ANM. However, quantitatively, the mixing was found to be
 a higher order effect and does not affect the pion dispersion relations at the
 leading order in density. Hence in the present paper we neglect
 $\p$-$\eta$ mixing.\\

 The plan of the paper is as follows. In section II we present the formalism,
 where we start with  PS coupling in subsection A. In II B we invoke non-linear
 field transformation \cite{Matsui82} and subsequently report results 
 involving PV coupling. In section III we present results using  
 recently developed chiral effective model in the context of nuclear many body
 problem \cite{Serot97,Hu07}. Finally, section IV presents the
 summary and conclusion. Detailed derivations for the Dirac part of the pion 
 self-energy for PS and PV couplings have been relegated to
 appendix A and B respectively.\\

\section{Formalism}

\subsection{Model with pseudoscalar $\p N$ interaction}

We start with the following interaction Lagrangian given by \cite{Matsui82},

\bwt
\bea
\mathcal{L} &=& \bar{\Psi}(i\g_\m \partial^\m - M)\Psi - \f{1}{2}g_\rho\bar{\Psi}\g_\m(\v{\t} \cdot \v{\Phi}^\m_\rho)\Psi + g_s\bar{\Psi}\Phi_s\Psi - g_\o\bar{\Psi}\g_\m\Phi^\m_\o\Psi - ig_\p\bar{\Psi}\g_5(\v{\t}\cdot\v{\Phi}_\p)\Psi \nn \\
&+& \f{1}{2}(\partial_\m\Phi_s\partial^\m\Phi_s - m^2_s\Phi^2_s) +
\f{1}{2}(\partial_\m \v{\Phi}_\p - g_\rho \v{\Phi}_{\rho\m}\times\v{\Phi}_\p)\cdot
(\partial^\m \v{\Phi}_\p - g_\rho \v{\Phi}^\m_{\rho}\times\v{\Phi}_\p)\nn \\
&-& \f{1}{2}m^2_\p\v{\Phi}^2_\p + \f{1}{2}g_{\phi\p}m_s\Phi_s\v{\Phi}^2_\p
- \f{1}{4}G_{\m\n}G^{\m\n} -\f{1}{4}\v{B}_{\m\n}\cdot\v{B}^{\m\n}
+ \f{1}{2}m^2_\o\Phi_{\o\m}\Phi^\m_\o + \f{1}{2}m^2_\rho\v{\Phi}_{\rho\m}\cdot\v{\Phi}^\m_\rho
\label{Lag00}
\eea
\ewt

where,

\bse
\bea
G_{\m\n} &=& \partial_\m \Phi_{\o\n} - \partial_\n \Phi_{\o\m} \\
\v{B}_{\m\n} &=&  \partial_\m \v{\Phi}_{\rho\n} - \partial_\n \v{\Phi}_{\rho\m}
- g_\rho \v{\Phi}_{\rho\m}\times\v{\Phi}_{\rho\n}.
\eea
\ese

 Here, $\Psi$, $\v{\Phi}_\p$, $\Phi_s$, $\v{\Phi}_\rho$ and $\Phi_\o$ 
 represents the nucleon, pion, sigma, rho and omega fields
 respectively and their masses are denoted by $M$, $m_\p$, $m_s$,
 $m_\rho$ and $m_\o$. This model successfully
 reproduces the saturation properties of nuclear matter and yields 
 accurate results for closed shell nuclei in the Dirac-Hartree 
 approximation \cite{Horowitz81}.\\

 It is to be noted that in Eq.(\ref{Lag00}) the pion-nucleon dynamics 
 is described by

\bea
{\cal L}^{PS}=-ig_\p\bar{\Psi}\g_5 \lt(\v{\t}\cdot
{\v{\Phi}}_\p\rt){\Psi} \label{Lps}
\eea

 where, $g_\p$ is the pion-nucleon coupling constant with 
 $\f{g^2_\p}{4\p}=12.6$ \cite{Engel96}. This apart, the interaction 
 Lagrangian of Eq.(\ref{Lag00}) also has another term involving the 
 coupling of pions with the scalar meson given by

\be
\mathcal{L}_s = \f{1}{2}g_{\phi\p} m_s \Phi_s \v{\Phi}^2_\p \label{Lsp}
\ee

\begin{figure}[htb]
\begin{center}
\includegraphics[scale=0.35,angle=0]{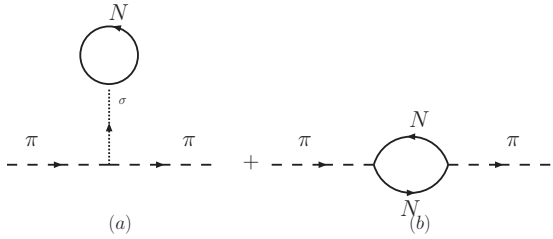}
\caption{Tadpole contribution to the pion self-energy}
\label{fig0}
\end{center}
\end{figure}

 Here, $g_{\phi\p}$ is the coupling constant of the scalar to pion
 field. The $\p$N scattering amplitude would now involve both nucleon
 and sigma meson in the intermediate state causing sensitive
 cancellation between the two that gives reasonable value of the
 $s$-wave scattering length \cite{Kapusta81} as mentioned
 before. At the self-energy level Eq.(\ref{Lps}) and (\ref{Lsp}) will 
 generate the exchange and the tadpole diagram  as shown in
 Fig.\ref{fig0}b and \ref{fig0}a.\\

%%%%%%%%%%%%%%%%%%%% TADPOLE DIAGRAM %%%%%%%%%%%%%%%%%%%%%%%%%%%%%%%
%%%%%%%%%%%%%%%%%%%%%%%%%%%%%%%%%%%%%%%%%%%%%%%%%%%%%%%%%%%%%%%%%%%%

 First we consider the tadpole diagram whose contribution to the self-energy
 is given by $\S^{TC} = - g_{\phi\p} m_s \phi_0$ where, 
 $\phi_0 = \f{g_s}{m^2_s}\rho^s$ and $\rho^s (= \rho^s_p +
 \rho^s_n)$. Here $\rho^s_i (i = p,n)$ represents scalar density
 given by

\bea
\rho^s_i=\frac{M^*_i}{2\pi^2} \left[E^*_i k_i - M^{* 2}_i \ln
\left ( \frac{E^*_i + k_i}{M^*_i}\right )\right ]. \label{scalard}
\eea

 The effective nucleon mass $M^*_i$ as appears in Eq.(\ref{scalard})
 can be determined from the following self-consistent condition \cite{Serot86}.

\bea M^*_i=M_i- \frac{g^2_s}{m^2_s} (\rho^s_p+\rho^s_n) \label{effectivem} \eea

 It is clear from Eq.(\ref{effectivem}) that $\Dt M^* = M_n-M_p = \Dt M$
 as the nucleon masses are modified by scalar mean field \cite{Serot86} which
 does not distinguish between $n$ and $p$. Here, for the moment we
 neglect explicit symmetry breaking ($n$-$p$ mass difference) {\em
 i.e.} $M^*_p=M^*_n=M^*$.\\

 It is to be noted that in the mean field theory (MFT), only
 Fig.(\ref{fig0}a), {\em i.e.} the tadpole diagram contributes, while 
 Fig.(\ref{fig0}b) is neglected. The origin of tachyonic mode can now
 easily be understood. The pion mass in matter due to the tadpole is
 given by \cite{Kapusta81}

\bea
m_\pi^{* 2}&=& m_\pi^2+\S^{TC} \nn \\
           &=& m_\pi^2-g_{\phi\pi} m_s \phi_0\nn\\
           &=& m_\pi^2-\frac{g_{\phi\pi} g_s}{m_s} (\rho_n^s+\rho_p^s)
\eea

 The second term of the last equation is quite large even at densities
 far below $\rho_0$ density {\em viz.} for $\rho \sim 0.1\rho_0$ 
 $m^{*2}_\p < 0$.

%%%%%%%%%%%%%%%%%%%%%%%%%%%%%% LOOP DIAGRAMS %%%%%%%%%%%%%%%%%%%%%%%%%%%%
\begin{figure}[htb]
\begin{center}
\includegraphics[scale=0.35,angle=0]{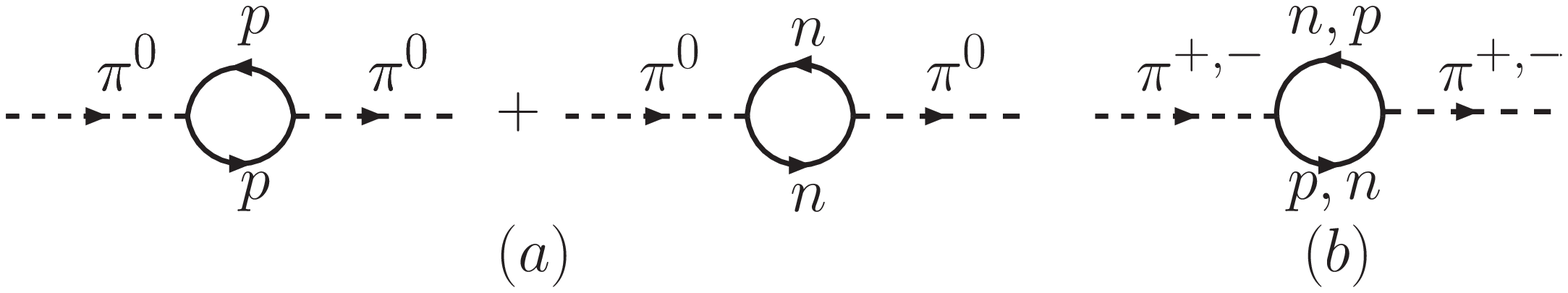}
\caption{(a) represents the one-loop self-energy diagram for $\p^0$,
and (b) represents the same for $\p^\pm$.}
\label{fig1}
\end{center}
\end{figure}
%%%%%%%%%%%%%%%%%%%%%%%%%%%%%%%%%%%%%%%%%%%%%%%%%%%%%%%%%%%%%%%%%%%%%%%%%%

 Fig.\ref{fig0}b would involve various combinations of $n$ and $p$
 depending upon the various charged states of pions as shown in 
 Fig.(\ref{fig1}a) and Fig.(\ref{fig1}b).

\bea
\S^*(q)&=& -i\int\f{d^4k}{(2\p)^4}\nn \\
& \times & Tr[\{i\G(q)\}iG_i(k+q)\{i\G(-q)\}iG_j(k)]\nn\\
& & \label{self0}
\eea

 where the subscript $i (j)$ denotes either $p$ (proton) or $n$ (neutron).
 $\G(q)$ is the vertex factor. $\G = -i\g_5$ or $- i\f{f_\p}{m_\p}\g_5\g_\m q^\m $
 for PS and PV coupling respectively. Explicitly,

%%%%%%%%%%%%%%%%%% NUCLEON PROPAGATORS %%%%%%%%%%%%%%%%%%%%%%

\bea G_i(k)=G^F_i(k)+G^D_i(k). \label{prop0} \eea where,

\bse
\label{prop1}
\bea
G^F_i(k)&=&\frac{k\!\!\!/+M^*_i}{k^2-M^{* 2}_i+i\z} \\
G^D_i(k)&=&\frac{i\pi(k\!\!\!/+M^*_i)}{E^*_i}\delta(k_0-E^*_i) \theta(k^F_i-|{\bf k}|)
\eea
\ese

 Here, $G^F_i(k)$ and $G^D_i(k)$ represent the free and the density dependent
 part of the propagator. In Eq.(\ref{prop1}) $k$ is the nucleon
 momentum; $k^F_i$ denotes the Fermi momentum and $M^*_i$ is the
 in-medium nucleon mass modified due to scalar mean field
 \cite{Serot86}. We, from now onward, use $k_p$ and $k_n$ to denote the
 proton and neutron Fermi momentum respectively.The nucleon energy is
 $E^*_i=\sqrt{M^{*2}_i + {\bf k}^2}$. \\

  Note that the total self-energy is given by $\S^*_{total}(q) = 
 \S^*(q) + \S^{TC}$. Using Eq.(\ref{prop0}) and Eq.(\ref{prop1}), the 
 expression for self-energy given in Eq.(\ref{self0}) takes the
 following form :

\bea
\S^*(q) &=& -ig^2\int\f{d^4k}{(2\p)^4}{\bf T} \nn \\
&=& \S^{*FF}(q) + \S^{*(FD+DF)}(q) + \S^{*DD}(q) \label{self1}
\eea

 Here $g$ is $g_\p$ $(f_\p/m_\p)$ for $PS$ (PV) coupling. For $\p^\pm$ the
 coupling constant $g_\p$ (or $f_\p$) gets replaced by
 $\sqrt{2}g_\p$. The values of the coupling constants $g_\p$ and
 $f_\p$ are determined experimentally from $\p$N and NN scattering
 data. ${\bf T}$ is the trace factor which consists of four parts :

\bea
{\bf T} = {\bf T}^{FF} + {\bf T}^{FD} + {\bf T}^{DF} + {\bf T}^{DD} \label{trace0}
\eea

\begin{figure}[htb]
\begin{center}
\includegraphics[scale=0.35,angle=0]{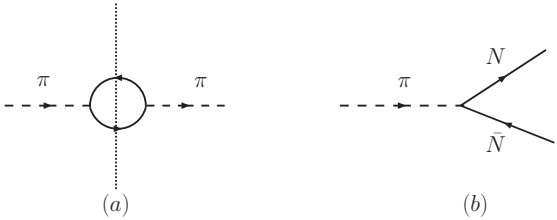}
\caption{(a) represents cutting of loop implied by the product of
two delta functions and (b) represents the decay of pion into
nucleon-antinucleon.}
\label{fig2}
\end{center}
\end{figure}

 Detailed expressions for ${\bf T}^{FF}$, ${\bf T}^{FD}$ and ${\bf
   T}^{DF}$ will
 be discussed later. Here the term ${\bf T}^{DD}$ contains the product of two
 delta functions ($G^D(k)G^D(k+q)$) which put both the loop-nucleons on shell
 implying the cut in the loop (Fig.\ref{fig2}a). This means that pion can decay
 into nucleon-antinucleon (Fig.\ref{fig2}b) pair which happens only in the high
 momentum limit {\em i.e} $q> 2k_{p,n}$ and also $q_0>2E^F_{p,n}$ where
 $E^F_{p,n}$ is the Fermi energy for proton (or neutron). Under this conditions
 only ${\bf T}^{DD}$ contributes to the self-energy. But in the present
 calculation, we investigate low momentum (of pion) collective excitations only
 \cite{Chin77}. Therefore ${\bf T}^{DD}$ ({\em i.e} $\S^{*DD}(q)$) is 
 neglected.\\

%%%%%%%%%%%%%%%%%%%%%%%%%%%%%% PION DECAY DIAGRAM %%%%%%%%%%%%%%%%%%%%%%%%%%%%%
%%%%%%%%%%%%%%%%%%%%%%%%%%%%%%%%%%%%%%%%%%%%%%%%%%%%%%%%%%%%%%%%%%%%%%%%%%%%%%%

 Thus, the pion self-energy can now be written as:

\bea
\S^*(q)&=& -ig^2\int\f{d^4k}{(2\p)^4}\lt[ {\bf T}^{FF} + \lt( {\bf T}^{FD}+{\bf T}^{DF}\rt) \rt]
\label{self3}
\eea

 The self-energies for different charged states of pion are calculated
 using the one-loop diagram shown in the Fig.(\ref{fig1}b). The first term of 
 Eq.(\ref{self3}) is same as the pion self-energy in vacuum with $M_i
 \ra M^*_i$. This part is divergent.

\bea
{\bf T}^{FF}_{PS} &= &2~Tr \lt[\g_5iG^F(k)\g_5iG^F(k+q) \rt]\nn \\
&=& -8\lt[ \f{M^{*2}-k\cdot(k+q)}{(k^2-M^{*2})\lt((k+q)^2-M^{*2} \rt)} \rt]\label{psself0}
\eea

 Here factor 2 that appears in ${\bf T}^{FF}_{PS}$, {\em i. e.} in 
 Eq.(\ref{psself0}), follows from isospin symmetry for $M_n =
 M_p$. The $FF$ part of self-energy  for PS coupling is calculated 
 from Eq.(\ref{self3}) by substituting ${\bf T}^{FF}_{PS}$ and
 it is denoted by $\S^{*FF}_{PS}(q)$.

\bea
\S^{*FF}_{PS}(q)&=& 8ig^2_\p\int\f{d^4k}{(2\p)^4}\nn \\
& \times & \lt[\f{M^{*2}-k\cdot(k+q)}{(k^2-M^{*2})\lt((k+q)^2-M^{*2} \rt)} \rt]\label{psff0}
\eea

 From Eq.(\ref{psff0}) it is observed that $\S^{*FF}_{PS}(q)$ is quadratically
 divergent. To eliminate these divergences we need to renormalize
 $\S^{*FF}_{PS}(q)$. Here we adopt the dimensional regularization
 \cite{Hooft73,Peskin,Cheng} technique to regularize
 $\S^{*FF}_{PS}(q)$ with the
 following results (details are discussed in appendix A).

%%%%%%%%%%%% RENORMALIZED FF PART OF PS COUPLING %%%%%%%%%%%%%%%%

\bwt
\bea \S^{*R}_{PS}(q,m_\p) &=& \f{g^2_\p}{2\p^2}\lt[-3(M^2-M^{*2})+\right.
\left.(q^2-m^2_\p)\lt(\f{1}{6}+\f{M^2}{m^2_\p}\rt) \right. \left. -
2M^{*2}\ln\lt(\f{M^*}{M}\rt)+\f{8M^2(M-M^*)^2}{(4M^2-m^2_\p)}\right.
\nn \\
&-& \left. \f{2M^{*2}\sqrt{4M^{*2}-q^2}}{q} \right. \left.
\tan^{-1}\lt(\f{q}{\sqrt{4M^{*2}-q^2}} \rt) \right. +\left.
\f{2M^2\sqrt{4M^2-m^2_\p}}{m_\p} \right. \left.
\tan^{-1}\lt(\f{m_\p}{\sqrt{4M^2-m^2_\p}} \rt) \right.
\nn \\
&+& \left. \lt( (M^2-M^{*2})+\f{m^2_\p(M-M^*)^2}{(4M^2-m^2_\p)} \right. \left.
+ \f{M^2}{m^2_\p}(q^2-m^2_\p) \rt)  \right. \left.
\f{8M^2}{m_\p\sqrt{4M^2-m^2_\p}} \right. \left.
\tan^{-1}\lt(\f{m_\p}{\sqrt{4M^2-m^2_\p} }\rt) \right.
\nn \\
&+& \left. \int^1_0 dx~3x(1-x)q^2 \right. \left. \ln\lt(
\f{M^{*2}-q^2x(1-x)}{M^2-m^2_\p x(1-x)} \rt)\right]
\label{apsrenorm1}
\eea
\ewt

 It is found that the result given in Eq.(\ref{apsrenorm1}) is finite and no
 divergences appear further. In the appropriate kinematic regime it might
 generate imaginary part:

%%%%%%%%%%%%%%%% IMAGINARY PART OF FF OF PS COUPLING %%%%%%%%%%%%%%%%%%%%

\bea
{\rm Im}~\S^{*FF}_{PS}(q) & = &
-\f{g^2_\p}{2\p^2}\int^1_0~dx~\lt(M^{*2}-3q^2x(1-x)\rt)\nn \\
&\times & {\rm Im}\lt[\ln\lt(M^{*2}-q^2x(1-x)-i\eta\rt)\rt] \nn \\
&=&-\f{g^2_\p}{4\p} \lt[q\sqrt{q^2-4M^{*2}}\rt]~\th\lt(q^2-4M^{*2}\rt)\nn \\
& &
\label{apsimaginary}
\eea

 If we consider that $(M^*-M)$ is small enough then the term
 $\ln[(M^{*2}-q^2x(1-x))/(M^2-m^2_\p x(1-x))]$ of Eq.(\ref{apsrenorm1}) can be
 approximated to $2\ln(M^*/M)$ and the last term of
 Eq.(\ref{apsrenorm1}) can be easily evaluated to give

\bea
\S^{*R}_{PS}(q,m_\p)\simeq - \tilde{\mathcal{C}} + \tilde{\mathcal{D}}q^2.
\label{psffapprox}
\eea

where,

\be\left.\begin{array}{ll} \tilde{\mathcal{C}}&= \f{g^2_\p}{2\p^2}
\lt[3(2M^2-M^{*2})+2M^{*2}\ln\lt(\f{M^*}{M}\rt)\rt] \\
& \\
\tilde{\mathcal{D}}&= \f{g^2_\p}{2\p^2} \lt[3\lt(\f{M}{m_\p} \rt)^2
\rt]\end{array}~ \right\}\ee

 The trace of (FD+DF) part for $\p^0$,

%%%%%%%%%%%%%% TRACE OF FD+DF OF PS COUPLING %%%%%%%%%%%%%%%%%%%%%%%%%%%%%%%%

\bea
T^{FD}_{PS}+T^{DF}_{PS} &=& Tr \lt[\g_5G^F_p(k+q)\g_5G^D_p(k) \rt. \nn \\
&+& \lt. \g_5G^D_p(k+q)\g_5G^F_p(k)\rt] + \lt[p \ra n\rt] \label{psfdpi00}
\eea

 and for $\p^{+(-)}$,

\bea
T^{FD}_{PS}+T^{DF}_{PS}&=&Tr \lt[\g_5G^F_{p(n)}(k+q)\g_5G^D_{n(p)}(k) \rt. \nn \\
&+& \lt. \g_5G^D_{p(n)}(k+q)\g_5G^F_{n(p)}(k) \rt] \label{psfdpm0}
\eea

%%%%%%%%%%%%%%%% SELF-ENERGY OF FD+DF OF PS COUPLING %%%%%%%%%%%%%%%%%%%%%%%%

 The $(FD+DF)$ part of the self-energy for $\p^0$ and $\p^\pm$ can be 
 written as

\bea
\S^{*0(FD+DF)}_{PS}(q)&=& -8g^2_\p\int \f{d^3k}{(2\p)^3E^*}{\bf A}_{PS}
\label{psfd01}\\
\S^{*\pm(FD+DF)}_{PS}(q)&=& -8g^2_\p\int \f{d^3k}{(2\p)^3E^*} [{\bf A}_{PS} \mp
{\bf B}_{PS}]
\nn  \\
& = & \S^{*0(FD+DF)}_{PS}(q)~\mp ~\dt\S^{*(FD+DF)}_{PS}(q),\nn \\
& & \label{psfdpm1}
\eea

 where,

\bea {\bf A}_{PS} & = & \lt[\f{(k\cdot q)^2}{q^4 - 4(k\cdot q)^2} \rt] (\th_p
+\th_n)
\label{aps}  \\
{\bf B}_{PS}& = &\f{1}{2}\lt[\f{q^2(k\cdot q)}{q^4-4(k\cdot q)^2} \rt] (\th_p -
\th_n) \label{bps} \eea

 with $\th_{p,n}=\th(k_{p,n}-|{\bf k}|)$. We restrict ourselves in the long
 wavelength limit {\em i.e.} when the pion momentum $({\bf q})$ is small
 compared to the Fermi momentum $(k_{p,n})$ of the system where the many body
 effects manifest strongly. In this case particle propagation can be understood
 in terms of  collective excitation \cite{Chin77} of the system which permits
 analytical solutions of the dispersion relations
 \cite{Chin77,Akdm03}. But in the
 short wavelength limit {\em i.e.} when the pion momentum $({\bf q})$ is much
 larger than the Fermi momentum $(k_{p,n})$, particle dispersion approaches to
 that of the free propagation. Note that for SNM ${\bf B}_{PS} = 0$ implying
 $\S^{*\pm(FD+DF)}_{PS} = \S^{*0(FD+DF)}_{PS}$.\\

 In the long wavelength limit we neglect the term $q^4$ compared to
 the term $4(k\cdot q)^2$ from the denominator of both ${\bf A}_{PS}$ and ${\bf
 B}_{PS}$ in Eqs.(\ref{aps}) and (\ref{bps}). Explicitly, after a straight
 forward calculation we get,

\bwt
\bea
\S^{*0(FD+DF)}_{PS}(q) &=& \f{g^2_\p}{2\p^2} \lt[\lt( k_p E^*_p -
\f{1}{2}M^{*2}\ln \lv\f{1+v_p}{1-v_p}\rv \rt) + \lt( k_n E^*_n - \f{1}{2}M^{*2}
\ln \lv\f{1+v_n}{1-v_n}\rv \rt) \rt] \label{psfd02}
\eea

and

\bea \dt\S^{*(FD+DF)}_{PS}(q) & = & \f{g^2_\p}{2\p^2}
\lt[\f{1}{2}E^*_p\ln\lv\f{c_0+v_p}{c_0-v_p}\rv - \right.
\left.\f{M^*}{\sqrt{c^2_0-1}}   \right. \left.
\tan^{-1}\lt(\f{k_p\sqrt{c^2_0-1}}{c_0M^*}\rt)\rt]\f{q^2}{|{\bf q}|}
\nn  \\
& - & \f{g^2_\p}{2\p^2} \lt[\f{1}{2}E^*_n\ln\lv\f{c_0+v_n}{c_0-v_n}\rv -
\right. \left.\f{M^*}{\sqrt{c^2_0-1}}   \right. \left.
\tan^{-1}\lt(\f{k_n\sqrt{c^2_0-1}}{c_0M^*}\rt)\rt]\f{q^2}{|{\bf q}|}
\label{psfdpm2}
\eea
\ewt

 where $v_{p,n}=k_{p,n}/E^*_{p,n}$, $E^*_{p,n}=\sqrt{M^{*2} + k^2_{p,n}}$ and
 $c_0=q_0/|{\bf q}|$. The approximate results of Eqs.(\ref{psfd02}) 
and(\ref{psfdpm2}) are given below.

\bea \S^{*0(FD+DF)}_{PS}(q) & \simeq & - \tilde{\mathcal{A}} -
\tilde{\mathcal{B}} -
\tilde{\mathcal{F}} + \tilde{\mathcal{G}} \label{psapprox} \\
\dt\S^{*(FD+DF)}_{PS}(q) & \simeq & \tilde{\mathcal{E}}
\f{q^2}{q_0}\label{psdapprox} \eea

where,

\be
\left.\begin{array}{ll} \tilde{\mathcal{A}} & =  \f{g^2_\p}{2\p^2}
\lt[\f{1}{3}\lt(\f{k^3_p}{E^{*3}_p} + \f{k^3_n}{E^{*3}_n}\rt) \rt]M^{*2} \\
& \\
\tilde{\mathcal{B}} & = \f{g^2_\p}{2\p^2} \lt[\f{1}{5}\lt(\f{k^5_p}{E^{*5}_p}
+ \f{k^5_n}{E^{*5}_n}\rt) \rt]M^{*2} \\
& \\
\tilde{\mathcal{F}} & = \f{g^2_\p}{2\p^2} \lt[\lt(\f{k_p}{E^*_p} +
\f{k_n}{E^*_n}\rt) \rt]M^{*2} \\
& \\
\tilde{\mathcal{G}} & = \f{g^2_\p}{2\p^2} \lt[k_pE^*_p + k_nE^*_n \rt] \\
& \\
\tilde{\mathcal{E}} & = \f{g^2_\p}{2\p^2} \lt[\f{1}{3}\lt(\f{k^3_p}{M^{*2}} -
\f{k^3_n}{M^{*2}}\rt) \rt]\end{array}~~~~~\right\} \ee

The self-energy for PS coupling:

\bea
\S^{*0,\pm}_{PS}(q) = \S^{*R}_{PS}(q,m_\p) + \S^{*0,\pm(FD+DF)}_{PS}(q) \label{pstotal}
\eea

 The dispersion relations can be found by solving Dyson-Schwinger equation.

\bea
q^2 - m^2_{\p^{0,\pm}} - (\S^{*0,\pm}(q) + \S^{TC}) = 0 \label{dyson}
\eea

 Here $m_{\p^{0,\pm}}$ are the masses of $\p^0$ and $\p^\pm$. The
 dispersion relations without the effect of Dirac sea for $\p^{0,\pm}$:

\bea
q^2_0 \simeq  m^{*2}_{\p^{0,\pm}}+ {\bf q}^2 \label{psdiswod}
\eea

 The effective masses without Dirac sea are

\bea
m^{*2}_{\p^0} \simeq \lt[ \O_{PS} + \S^{TC} + m^2_{\p^0}\rt] \nn \\
~~~{\rm and} ~~ \nn \\
m^{*2}_{\p^\pm} \simeq \lt[ \f{\O_{PS} + \S^{TC} + m^2_{\p^\pm}}{1 \mp \dt\O_{PS}} \rt]
\label{pseffectivemass}
\eea

where,

\be
\left.\begin{array}{ll} \O_{PS} &= \tilde{\mathcal{G}} - \tilde{\mathcal{A}}
- \tilde{\mathcal{B}} - \tilde{\mathcal{F}} \\
& \\
\dt\O_{PS} &= \f{\tilde{\mathcal{E}}}{ \sqrt{ \O_{PS} + \S^{TC} + m^2_{\p^\pm} } }
\end{array}~~~~~\right\}\label{psdlambda}
\ee

 Now we presents the dispersion relations for $\p^{0,\pm}$ with the
 effect of Dirac sea.

\bea
q^2_0 \simeq  m^{*2}_{\p^{0,\pm}}+ {\bf q}^2 \label{psdiswd}
\eea

 The effective masses $(m^{*2}_{\p^{0,\pm}})$ with Dirac sea for different
 charged states of pion are given by

\bea
m^{*2}_{\p^0} \simeq \lt[(\L_{PS} - m^2_{\p^0})/\tilde{\mathcal{D}}\rt] \nn \\
~~ {\rm and} ~~~ \nn \\
 m^{*2}_{\p^\pm} \simeq
\lt[ \f{(\L_{PS} - m^2_{\p^\pm})} {(1 \mp \dt\L_{PS})\tilde{\mathcal{D}} }\rt]
\label{pseffectivemass0}
\eea

where,

\be\left.\begin{array}{ll} \L_{PS} &= \tilde{\mathcal{C}} - \O_{PS} - \S^{TC} \\
& \\
\dt\L_{PS} &= \f{\tilde{\mathcal{E}}}{\sqrt{(\L_{PS} - m^2_{\p^\pm})\tilde{\mathcal{D}} }}
\end{array}~~~~~\right\}\label{psdlambda0}\ee

 The PS coupling the asymmetry driven mass splitting is
 of ${\mathcal O(k^3_{p(n)}/M^{*2})}$. The terms $\dt\L_{PS}$ and
 $\dt\O_{PS}$ are non-vanishing in ANM and responsible for the pion
 mass splitting.\\

\begin{figure}[htb]
\begin{center}
\includegraphics[scale=0.3,angle=0]{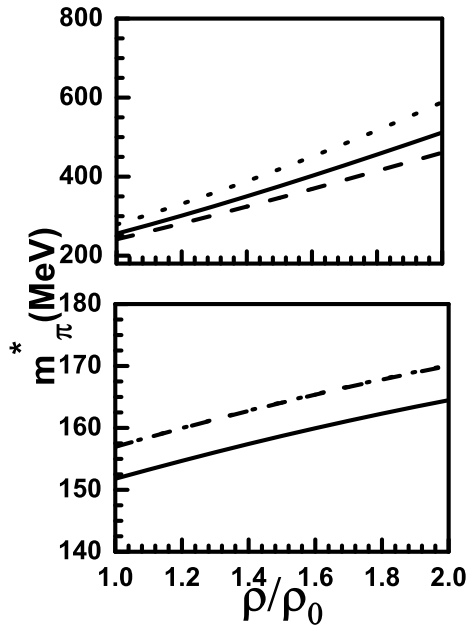}
\caption{Nuclear density $(\rho)$ dependent effective pion masses for PS
coupling at $\a=0.2$. The dotted, dashed and solid curves representing
$\p^-$, $\p^+$ and $\p^0$ without (upper panel) and with (lower panel) the 
Dirac sea effect.} \label{psfig1}
\end{center}
\end{figure}

\begin{figure}[htb]
\begin{center}
\includegraphics[scale=0.3,angle=0]{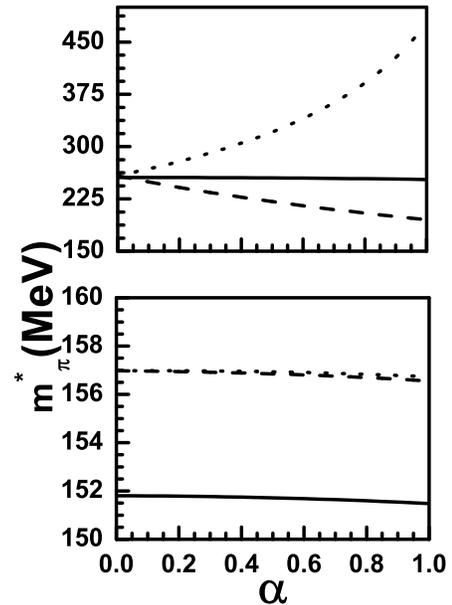}
\caption{Asymmetry parameter $(\a)$ dependent effective pion masses 
for $\p^0$ (solid curve), $\p^+$ (dashed curve) and $\p^-$ (dotted
curve) in ANM at NNM density for PS coupling.
The upper and lower panel represents effective pion masses without and with 
vacuum correction.} \label{psfig2}
\end{center}
\end{figure}

 In Fig.\ref{psfig1} and \ref{psfig2} we present the density ($\rho$) and
 asymmetry parameter ($\a$) dependent effective masses for the various charged
 states of pion. In the top panel we present the results without
 vacuum correction (Dirac sea). Here we include both the tadpole and
 $n$-$n$ loop.

%%%%%%%%%%%%%%%%%%%%%%%%%%% PS RHO %%%%%%%%%%%%%%%%%%%%%%%%%%%%%%%%%%%%%%%%%
%%%%%%%%%%%%%%%%%%%%%%%%%%%%%%%%%%%%%%%%%%%%%%%%%%%%%%%%%%%%%%

 It is evident that the inclusion of (\ref{fig0}b) removes the tachyonic mode
 but gives rise to effective pion masses which are unrealistically
 large as discussed by Kaputa \cite{Kapusta81} as shown in the top
 panel of Fig.\ref{psfig1}. \\

%%%%%%%%%%%%%%%%%%%%%%%%%%%%%%%%%%%%%%%%%%%%%%%%%%%%%%%%%%%%%%%
\begin{table}
\caption{This table presents the effective pion masses including the
  tadpole contribution to the self-energy in $PS$ coupling. Kapusta 
  corresponds to ref. \cite{Kapusta81} and BDM corresponds to the
  present calculation.}
\begin{ruledtabular}
\label{masses}
\begin{tabular}{ccc}
           & $m^{*2}_{\p^0}$         & $m^{*2}_{\p^\pm}$ \\ \hline
           &                         &                          \\
  MFT      & $m^2_{\p^0} + \S^{TC}$  & $m^2_{\p^\pm} + \S^{TC}$ \\
           &                         &                          \\
  Kapusta  & $\O_{PS} + (\S^{TC} + m^2_{\p^0})$
  & $\f{\O_{PS} + (\S^{TC} + m^2_{\p^\pm}) }{1\mp \dt\O_{PS}}$ \\
           &                         &                         \\
  BDM  & $[\tilde{\mathcal{C}} - (\O_{PS} + \S^{TC} + m^2_{\p^0})] /\tilde{\mathcal{D}}$
  & $\f{\tilde{\mathcal{C}} - (\O_{PS} + \S^{TC} + m^2_{\p^\pm})}
  {(1 \mp \dt\L_{PS})\tilde{\mathcal{D}} }$
\end{tabular}
\end{ruledtabular}
\end{table}

%%%%%%%%%%%%%%%%%%%% PS ALFA %%%%%%%%%%%%%%%%%%%%%%%%%%%%%%%%%%%%%%%
%%%%%%%%%%%%%%%%%%%%%%%%%%%%%%%%%%%%%%%%%%%%%%%%%%%%%%%%%%%%%%%%%%%%%

 It is to be noted that the inclusion of the vacuum part reduces the
 effective pion masses and gives reasonable value for the density
 dependent pion masses in matter at NNM density. The reason for this
 could be understood from the Table~\ref{masses} which enumerates 
 expressions for the effective pion masses that we obtain in three
 different cases. The top row represents effective pion masses for the 
 case considered in \cite{Serot79} which gives rise to the tachyonic
 mode, the second row corresponds to the case discussed by Kapusta 
 \cite{Kapusta81} and in the last row we present results of the present work
 as by BDM. The presence of the additional term $\tilde{\mathcal{D}}$ 
 somewhat tames the dispersion curve bringing the masses down compared
 to \cite{Kapusta81}. This can be noted that at the MFT level $\S^{TC}$
 involves sum of the scalar densities $\rho^s_n$ and $\rho^s_p$. 
 Therefore, in MFT, as expected, the masses are insensitive to
 asymmetry parameter $\a$.

%%%%%%%%%%%%%%%%%%%%% PS DISPERSION %%%%%%%%%%%%%%%%%%%%%%%%%%%%
\begin{figure}[htb]
\begin{center}
\includegraphics[scale=0.3,angle=0]{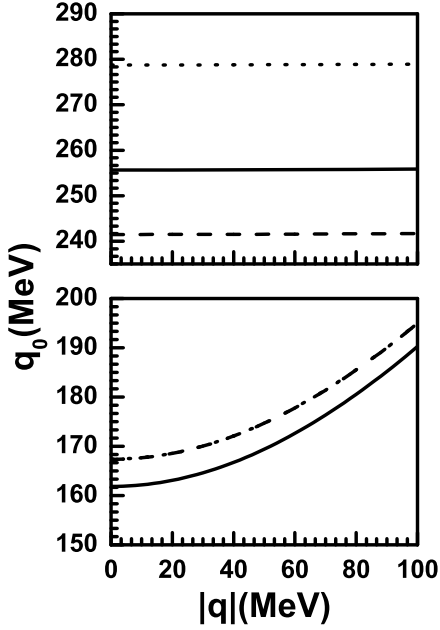}
\caption{The dispersion relations of $\p^0$ (solid curve), $\p^+$ 
 (dashed curve) and $\p^-$ (dotted curve) for PS coupling at 
 $\rho = 0.17fm^{-3}$ and $\a = 0.2$. The upper and lower panel
 representing pion dispersions without and with the Dirac sea.} 
\label{psfig3}
\end{center}
\end{figure}
%%%%%%%%%%%%%%%%%%%%%%%%%%%%%%%%%%%%%%%%%%%%%%%%%%%%%%%%%%%%%%%%

 For Pb-like nuclei ($\a = 0.2$), $\Dt m_{\p^0} = 16.8 MeV$, $\Dt
 m_{\p^+} = 17.37 MeV$ and $\Dt m_{\p^-} = 17.41 MeV$ with vacuum correction.\\

 The dispersion relation for various charged states of pion are shown
 in the Fig.\ref{psfig3} where upper and lower panel present the
 dispersion curves without and with the Dirac sea including the
 tadpole contribution. In the presence of Dirac sea $\p^+$ and $\p^-$ 
 dispersion curves are indistinguishable. \\

\subsection{Renormalizible model with pseudovector $\p N$ coupling}

 To obtain the pseudovector representation from the starting
 Lagrangian given by Eq.(\ref{Lag00}), one gives following nonlinear 
 chiral transformation \cite{Matsui82},

\bse
\bea
\Psi &=& \lt[ \f{1 - i\g_5\v{\t}\cdot\v{\xi}}{\sqrt{1 + \v{\xi}^2}} \rt] \Psi^\prime \label{ncsi}\\
\v{\xi} &=& \lt( \f{f_\p}{m_\p}\rt)\v{\Phi}^\prime_\p \nn \\
&=& g_\p \v{\Phi}_\p/\lt[M - g_s \Phi_s + \sqrt{(M - g_s \Phi_s)^2 + g^2_\p \v{\Phi}^2_\p} \rt] \nn \\
& & \label{ncpi}\\
g_s\Phi^\prime_s &=& M - \sqrt{(M - g_s \Phi_s)^2 + g^2_\p \v{\Phi}^2_\p}\label{ncphi}
\eea
\ese

 The last two equations (\ref{ncpi}) and (\ref{ncphi}) are used to
 express the old fields $\Phi_s$ and $\v{\Phi}_\p$ in terms of new
 fields $\Phi^\prime_s$ and $\v{\Phi}^\prime_\p$.

\bse
\bea
\v{\Phi}_\p &=& \lt[ \f{1-2(f_\p/m_\p) \Phi^\prime_s}{1 + (f_\p/m_\p)^2 \v{\Phi}^{\prime 2}_\p}\rt] \v{\Phi}^\prime_\p \\
\Phi_s &=& \f{(1- (f_\p/m_\p)^2\v{\Phi}^{\prime 2}_\p)\Phi^\prime_s +
(g_\p/g_s)(f_\p/m_\p)\v{\Phi}^{\prime 2}_\p}{1 + (f_\p/m_\p)^2\v{\Phi}^{\prime 2}_\p}\nn \\
& &
\eea
\ese

\bwt
The transformed Lagrangian is
\bea
\mathcal{L}^\prime &=& \bar{\Psi}^\prime (i\g_\m \partial^\m -M)\Psi^\prime - \f{1}{2}g_\rho \bar{\Psi}^\prime \g_\m (\v{\t}\cdot\v{\Phi}^\m_\rho )\Psi^\prime +
g_s \bar{\Psi}^\prime \Phi^\prime_s \Psi^\prime - g_\o \bar{\Psi}^\prime \g_\m \Phi^\m_\rho \Psi^\prime
 \nn \\
%-----------------------------%----------------------------------------------
&+& \f{1}{2}(\partial_\m \Phi_s\partial^\m\Phi_s - m^2_s \Phi^2_s) +
\f{1}{2}(\partial_\m \v{\Phi}_\p - g_\rho \v{\Phi}_{\rho\m}\times \v{\Phi}_\p)\cdot
(\partial^\m \v{\Phi}_\p - g_\rho \v{\Phi}^\m_\rho\times \v{\Phi}_\p) \nn \\
%-----------------------------%----------------------------------------------
&-& \f{1}{2}m^2_\p \v{\Phi}^2_\p + \f{1}{2}g_{\phi\p}m_s \Phi_s \v{\Phi}^2_\p
- \f{1}{4}G_{\m\n}G^{\m\n} - \f{1}{4}\v{B}_{\m\n}\cdot\v{B}^{\m\n}
+ \f{1}{2}m^2_\o \Phi_{\o\m}\Phi^\m_\o
+ \f{1}{2}m^2_\rho \v{\Phi}_{\rho\m}\cdot\v{\Phi}^\m_\rho \nn \\
%----------------------------%-----------------------------------------------
&-&\f{(f_\p/m_\p)^2}{1+(f_\p/m_\p)^2\v{\Phi}^{\prime 2}_\p}\bar{\Psi}^\prime \g_\m(\v{\t}\cdot\v{\Phi}^\prime_\p)
\times (\partial^\m\v{\Phi}^\prime_\p - g_\rho \v{\Phi}^\m_\rho\times\v{\Phi}^\prime_\p)\Psi^\prime
-\f{(f_\p/m_\p)}{1+(f_\p/m_\p)^2\v{\Phi}^{\prime 2}_\p}\bar{\Psi}^\prime \g_5 \g_\m \v{\t}\cdot
(\partial^\m\v{\Phi}^\prime_\p - g_\rho \v{\Phi}^\m_\rho\times\v{\Phi}^\prime_\p)\Psi^\prime \nn \\
& & \label{TLag}
\eea
\ewt

 We see from the Eq.~{\ref{TLag}} that the $\pi$-$N$ PS coupling has 
 disappeared and instead the pion-nucleon dynamics is now governed by
 the last term of the above mentioned equation. At the
 leading order one obtains the usual PV coupling represented by,

\bea
{\cal L}^{PV} = -\f{f_\p}{m_\p}\bar{\Psi}^\prime
\g_5\g^\m{\partial}_\m \lt(\v{\t}\cdot \v{\Phi}^\prime_\p\rt)\Psi^\prime
\label{Lpv}
\eea

 Here $f_\p$ is the pseudovector coupling constant and
 $\f{f^2_\p}{4\p} = 0.08$ \cite{Ericson88}.
 First we discuss the FF part where the trace factor is given by

%%%%%%%%%%%%%%%% TRACE OF FF OF PV COUPLING %%%%%%%%%%%%%%%%%%%%%%%%%%

\bea
{\bf T}^{FF}_{PV} &=& -2~Tr[\g_5\g^\m q_\m iG^F(k)\g_5\g^\n
q_\n iG^F(k+q)] \nn \\
&=& -8~\lt[\f{M^{*2}q^2+k\cdot (k+q)q^2-2(k\cdot q)(k+q)\cdot q}
{(k^2-M^{*2})((k+q)^2-M^{*2})}\rt]\nn \\
& & \label{pvself0}
\eea

 Now the $FF$ part of the self-energy for $PV$ coupling is denoted by
$\S^{*FF}_{PV}(q)$. From Eq.(\ref{self3}) and Eq.(\ref{pvself0}) we get,

%%%%%%%%%%%%%%%%%% SELF-ENERGY OF FF OF PV COUPLING %%%%%%%%%%%%%%%%%%%%%

\bea
\S^{*FF}_{PV}(q)&=& 8i\lt(\f{f_\p}{m_\p}\rt)^2\int\f{d^4k}{(2\p)^4}\nn \\
& \times & \lt[\f{M^{*2}q^2+k\cdot (k+q)q^2-2(k\cdot q)(k+q)\cdot q}
{(k^2-M^{*2})((k+q)^2-M^{*2})}\rt]\nn \\
& & \label{pvff0}
\eea

 Direct power counting shows that the term $\S^{*FF}_{PV}(q)$ is divergent.
 The appropriate renormalization scheme for the present model has been 
 developed in ref\cite{Matsui82}. We first consider a simple
 subtraction scheme described in Appendix B to obtain

\bea
\S^{*R}_{PV}(q)&=& \f{q^2}{2\p^2}\lt(\f{f_\p}{m_\p}\rt)^2 \nn \\
& \times & \lt[2M^{*2}\int^1_0dx~ \ln\lt(\f{M^{*2}-q^2x(1-x)}{M^{*2}-m^2_{\p}x(1-x)}\rt)\rt]\nn \\
\label{apvrenorm1}
\eea

 Now $\S^{*R}_{PV}(q)$ can be approximated to

\bea \S^{*R}_{PV}(q)\simeq {\cal C}q^2 - {\cal D}q^4 \label{apvffapprox} \eea

where,

\be \left. \begin{array}{ll} {\cal C} & = \lt(\f{f_\p
M^*}{m_\p\p}\rt)^2\lt[\f{m^2_\p}{6M^{*2}}\rt] \\
& \\
{\cal D} & = \lt(\f{f_\p M^*}{m_\p\p}\rt)^2 \lt[\f{1}{6M^{*2}}\rt]
\end{array}~\right\} \label{Dpipi} \ee

%%%%%%%%%%%%%%%%%%%%%%%%%  MATSUI & SEROT %%%%%%%%%%%%%%%%%%%%%%%%%%%%%%%%%%%

 On the other hand borrowing results from \cite{Matsui82} one has,

\bea \S^{*R}_{PV}(q)\simeq {\cal C}^\prime + {\cal D}^\prime q^2 \label{matsui0} \eea

 where,

\be \left. \begin{array}{ll} {\cal C}^\prime & = \lt(\f{f_\p}{m_\p\p}\rt)^2
\lt[\f{4}{3}M(M-M^*)m^2_\p \rt] \\
& \\
{\cal D}^\prime & = \lt(\f{f_\p}{m_\p\p}\rt)^2 \lt[2M^{*2}\ln\lt(M^*/M \rt) \rt]
\end{array}~\right\} \label{matsui1} \ee

 It might be mentioned, although ${\cal C, D}$ are different from 
 ${\cal C^\prime, D^\prime}$, their effect on the effective pion
 masses and corresponding dispersion relations are
 found to be marginal as we discuss later.

 The FF part can also develop imaginary part as given by

\bea
{\rm Im}~\S^{*FF}_{PV}(q)&=& -\lt(\f{f_\p}{m_\p}\rt)^2\nn \\
& \times & \lt[\f{q}{\p}2M^{*2}\sqrt{q^2-4M^{*2}} \rt]~\th\lt(q^2-4M^{*2}\rt) \nn \\
& & \label{apvimaginary}
\eea

 It is observed from Eq.(\ref{apvimaginary}) that ${\rm
   Im}~\S^{*FF}_{PV}(q)$ is non-vanishing only if $q^2>4M^{*2}$.

%%%%%%%%%%%%%%%%% FD + DF PART OF PV COUPLING %%%%%%%%%%%%%%%%%%%%%%%%%%%%%%%%%

 The trace of the (FD+DF) part for $\p^0$,

\bea
T^{FD}_{PV}+T^{DF}_{PV}&=& Tr \lt[\g_5q\!\!\!/G^F_p(k+q)\g_5q\!\!\!/G^D_p(k)\rt. \nn \\
&+& \lt. \g_5q\!\!\!/G^D_p(k+q)\g_5q\!\!\!/G^F_p(k)\rt]+[p\ra n] \nn \\
& & \label{pvfdpi00}
\eea

and for $\p^{+(-)}$,

\bea
T^{FD}_{PV}+T^{DF}_{PV}&=& Tr \lt[\g_5q\!\!\!/G^F_{p(n)}(k+q)\g_5q\!\!\!/G^D_{n(p)}(k) \rt. \nn \\
&+& \lt. \g_5q\!\!\!/G^D_{p(n)}(k+q)\g_5q\!\!\!/G^F_{n(p)}(k)\rt]\label{pvfdpm0}
\eea

 In pure neutron (or proton) matter one of the terms of
 Eq.(\ref{pvfdpm0}) {\em viz} $G^D_{p(n)} = 0$ for the charged pion
 states. The same argument holds true
 for the neutral pion where only two terms would contribute which can be
 observed from  Eq.(\ref{pvfdpi00}). In case of pure neutron (or
 proton) matter $p(n)$ appears as the intermediate state. Now the 
 $(FD+DF)$ part of the self-energy for $\p^0$ and $\p^\pm$ can be written as

\bea
\S^{*0(FD+DF)}_{PV}(q) &=& -8\lt(\f{f_\p}{m_\p}\rt)^2
\int\f{d^3k}{(2\p)^3E^*}{\bf A}_{PV}\label{pvfd01}\\
\S^{*\pm(FD+DF)}_{PV}(q) &=& -8\lt(\f{f_\p}{m_\p}\rt)^2
\int\f{d^3k}{(2\p)^3E^*}[{\bf A}_{PV}\mp{\bf B}_{PV}] \nn \\
&=& \S^{*0(FD+DF)}_{PV}(q)\mp \dt\S^{*(FD+DF)}_{PV}(q) \nn \\
& & \label{pvfdpm1}
\eea

 where

\bea
{\bf A}_{PV} &=&\lt[\f{M^{*2}q^4}{q^4-4(k\cdot
q)^2}\rt](\th_p+\th_n)\label{apv}\\
{\bf B}_{PV} &=&\f{1}{2}\lt[1+\f{4M^{*2}q^2}{q^4-4(k\cdot q)^2}\rt]
(k\cdot q)(\th_p-\th_n)\label{bpv}
\eea

 In the long wavelength limit considering collective excitations near the
 Fermi surface, $(FD+DF)$ part of the pion-self energy can be
 evaluated analytically. In this case we can neglect the term $q^4$ 
 compared to the term $4(k\cdot q)^2$ from the denominator of ${\bf
 A}_{PV}$ and ${\bf B}_{PV}$ in Eqs. (\ref{apv}) and (\ref{bpv}). This
 is called hard nucleon loop (HNL) approximation
 \cite{Akdm03}. Explicitly, after a straight forward calculation, we get,

\bwt
\bea
\S^{*0(FD+DF)}_{PV}(q) &=& \f{1}{2}\lt(\f{f_\p M^*}{m_\p\p}\rt)^2
\lt[\lt(\ln\lv\f{1+v_p}{1-v_p}\rv-c_0\ln\lv\f{c_0+v_p }{c_0-v_p}\rv\rt) +
\lt(\ln\lv\f{1+v_n}{1-v_n}\rv-c_0\ln\lv\f{c_0+v_n }{c_0-v_n}\rv\rt)
\rt]\label{pvfd02}
\eea

and

\bea
\dt\S^{*(FD+DF)}_{PV}(q) &=& \lt(\f{f_\p}{m_\p\p}\rt)^2\lt[
\f{2}{3}k^3_p q_0 -\f{M^{*2}q^2}{|{\bf
q}|}\lt(E^*_p\ln\lv\f{c_0+v_p}{c_0-v_p}\rv
-\f{2M^*}{\sqrt{c^2_0-1}}\tan^{-1}\f{k_p\sqrt{c^2_0-1}}{c_0M^*}\rt)\rt]\nn
\\
&-& \lt(\f{f_\p}{m_\p\p}\rt)^2\lt[ \f{2}{3}k^3_n q_0
-\f{M^{*2}q^2}{|{\bf q}|}\lt(E^*_n\ln\lv\f{c_0+v_n}{c_0-v_n}\rv
-\f{2M^*}{\sqrt{c^2_0-1}}\tan^{-1}\f{k_n\sqrt{c^2_0-1}}{c_0M^*}\rt)\rt]
\label{pvfdpm2}
\eea
\ewt

 The approximate results of Eqs.(\ref{pvfd02}) and (\ref{pvfdpm2}) are

\bea
\S^{*0(FD+DF)}_{PV}(q) &\simeq & {\cal A} \f{q^4}{q^2_0} +
{\cal B} q^2 \label{pvapprox} \\
\dt\S^{*(FD+DF)}_{PV}(q) & \simeq & {\cal E}q_0 \label{pvdapprox}
\eea

where,

\be
\left. \begin{array}{ll} {\cal A} &= \lt(\f{f_\p M^*}{m_\p\p} \rt)^2
\lt[\f{1}{3}\lt(\f{k^3_P}{E^{*3}_p} + \f{k^3_n}{E^{*3}_n} \rt)\rt]\\
& \\
{\cal B} &= \lt(\f{f_\p M^*}{m_\p\p} \rt)^2 \lt[\f{1}{5}\lt(\f{k^5_P}{E^{*5}_p}
+ \f{k^5_n}{E^{*5}_n}\rt)\rt]\\
& \\
{\cal E} &= \lt(\f{f_\p M^*}{m_\p\p} \rt)^2 \lt[\f{2}{5}\lt(\f{k^5_p}{M^{*4}} -
\f{k^5_n}{M^{*4}}\rt)\rt] \end{array} ~ \right\} \label{Epipi}
\ee

 The total pion self-energy for $PV$ coupling is

\bea \S^{*0,\pm}_{PV}(q) = \S^{*R}_{PV}(q) + \S^{*0,\pm
(FD+DF)}_{PV}(q) \label{pvtotal} \eea

 The approximate dispersion relations and the effective pion masses of 
 different charged states in ANM without and with the Dirac sea effect
 are presented  below.\\

 The dispersion relations for $\p^{0,\pm}$ without the effect of Dirac
 sea are as follows,

\bea
q^2_0 \simeq m^{*2}_{\p^{0,\pm}} + \g_{\p\p}{\bf q}^2 + \lt[\f{\g^2_{\p\p}}{4} +
\a_{\p\p}\rt]\f{{\bf q}^4}{m^{*2}_{\p^{0,\pm}}}\label{pvdiswd}
\eea

where $m^{*2}_{\p^{0,\pm}}$ is the effective pion masses without Dirac
sea effect:

\bea
m^{*2}_{\p^0} \simeq \f{m^2_{\p^0}}{1-\O_{PV}} ~ {\rm and} ~
m^{*2}_{\p^{\pm}}\simeq \f{m^2_{\p^{\pm}}}{1-(\O_{PV} \pm {\dt\O}_{PV})}
\label{pveffectivemasseswd}
\eea

where,

\be\left.\begin{array}{ll}
\O_{PV} &= {\cal A + B} \\
& \\
{\dt\O}_{PV}&=\f{{\cal E}}{m_{\p^{\pm}}}\\
& \\
\g_{\p\p} &= 1 - \f{\O_{PV}}{1-\O_{PV}} + \f{{\cal B}}{1-\O_{PV}}\\
& \\
\a_{\p\p}&= \f{{\cal A}}{1-\O_{PV}}\end{array}~\right\}\label{pvalpha}
\ee

 These results are the same as that of \cite{Biswas06} with some notational
 difference such as $\O_{PV} \ra \O^2_{\p\p}$, $\dt\O_{PV} \ra \dt\O^2_{\p\p}$,
 $\O_{PV}/(1 - \O_{PV}) \ra \chi_{\p\p}$ and $\mathcal{B}/(1 - \O_{PV}) \ra
 \b_{\p\p}$. In (\ref{pveffectivemasses}) and (\ref{pveffectivemasseswd}),
 ${\dt\L}_{PV}$ and ${\dt\O}_{PV}$ are responsible for the asymmetry parameter
 $(\a= \f{\rho_n -\rho_p}{\rho_n + \rho_p})$ dependent mass splitting, where
 $\rho_n$ and $\rho_p$ are the neutron and proton density respectively. Clearly
 for SNM  ${\dt\L}_{PV}$ and ${\dt\O}_{PV}$ vanish.\\

 The dispersion relations for $\p^{0,\pm}$ including the effect of
 Dirac sea are given by,

\bea
q^2_0 & \simeq & m^{*2}_{\p^{0,\pm}} + \lt[\g_{\p\p}+2m^{*2}_{\p^{0,\pm}}\dt_{\p\p}\rt]{\bf q}^2 \nn \\
&+& \lt[\f{\g^2_{\p\p}}{4} + \a_{\p\p} - \dt_{\p\p} \lt(m^{*2}_{\p^{0,\pm}} -
2\g_{\p\p}\rt) \rt]\f{{\bf q}^4}{m^{*2}_{\p^{0,\pm}}} \nn \\
& & \label{pvdis}
\eea

 The effective masses $(m^*_{\p})$ of different charged states of pion
 are found from Eq.({\ref{pvdis}) in the limit $|{\bf q}|=0$.

\bea
m^{*2}_{\p^0} \simeq \f{m^2_{\p^0}}{1-\L_{PV}} ~{\rm and} ~
m^{*2}_{\p^{\pm}}\simeq \f{m^2_{\p^{\pm}}}{1-(\L_{PV} \pm {\dt\L}_{PV})}
\label{pveffectivemasses}
\eea

where,

\be
\left.\begin{array}{ll}
\L_{PV} &= {\cal A + B + C}\\
& \\
{\dt\L}_{PV} &= \f{{\cal E}}{m_{\p^\pm}} \\
& \\
 \g_{\p\p} &=1 - \f{\L_{PV}}{1-\L_{PV}} + \f{{\cal
B}}{1-\L_{PV}} + \f{{\cal C}}{1-\L_{PV}} \\
& \\
\a_{\p\p} &= \f{{\cal A}}{1-\L_{PV}} \\
& \\
\dt_{\p\p} &= \f{{\cal D}}{1-\L_{PV}}\end{array}~\right\} \label{pvdelta}
\ee

 This is to be noted that, if one use Eq.(\ref{matsui0}) instead of 
 Eq.(\ref{apvffapprox});  $m^2_{\p^{0,\pm}}$ and ${\cal C}$ will be
 replaced by $(m^2_{\p^{0,\pm}} + {\cal C}^\prime)$
 and ${\cal D}^\prime$ respectively. $\dt_{\p\p}$ will vanish. 
 Numerically,as mentioned before, Eq.(\ref{apvffapprox}) and 
 Eq.(\ref{matsui0}) give results very close to each other.
 Clearly from Eq.(\ref{Epipi}), ${\cal E}$ indicates that the asymmetry driven
 mass splitting is of ${\cal O}(k^5_{p(n)}/M^{*4})$ for $PV$ coupling.\\

%%%%%%%%%%%%%%%%%%%%%%%%%%%%%%%%%%%%%%%%%%%%%%%%%%%%%%%%%%
\begin{figure}[htb]
\begin{center}
\includegraphics[scale=0.3,angle=0]{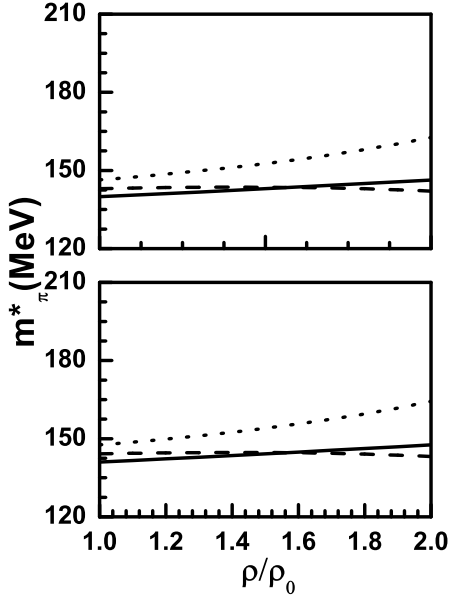}
\caption{Effective masses (for PV coupling) of $\p^0$ (solid curve), 
$\p^+$ (dashed curve) and $\p^-$ (dotted curve) are represented
  without Dirac sea (upper panel) and with Dirac sea
(lower panel) at $\a = 0.2$.}
\label{pvfig1}
\end{center}
\end{figure}
%%%%%%%%%%%%%%%%%%%%%%%%%%%%%%%%%%%%%%%%%%%%%%%%%%%%%%%%%

 To quote typical values of the pion mass shifts for PV coupling at
 normal nuclear
 density $(\rho_0 = 0.17 fm^{-3})$ for Pb-like nuclei which are 
 $\Dt m_{\p^0} = 6.07 MeV$, $\Dt m_{\p^+} = 4.6 MeV$ and 
 $\Dt m_{\p^-} = 8.02 MeV$ with vacuum correction and the
 corresponding values are $4.95 MeV$, $3.47 MeV$ and $6.82 MeV$ 
 without vacuum correction.\\

 In Fig.\ref{pvfig1} we show results for the density dependence of 
 effective pion masses for various charge states at $\a = 0.2$. 
 It is observed that the $\p^-$ mass increases in matter while 
 $\p^+$ decreases at higher density. The mass splitting is quite significant
 even at density $\rho \gtrsim 1.5\rho_0$. In the lower panel we
 present results with vacuum  corrections. Evidently the effect of
 vacuum corrections is found to be small. \\

%%%%%%%%%%%%%%%%%%%%%%%%%%%%%%
\begin{figure}[htb]
\begin{center}
\includegraphics[scale=0.3,angle=0]{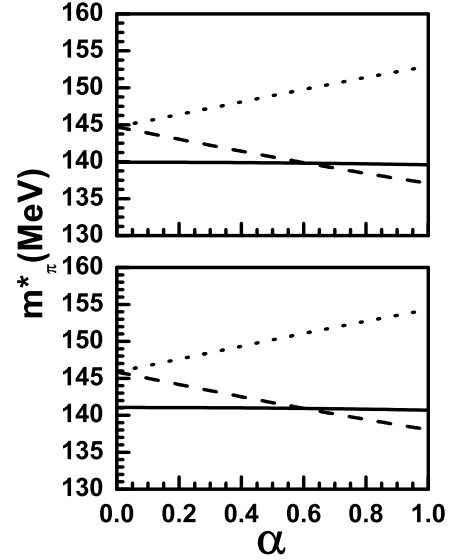}
\caption{Asymmetry parameter $(\a)$ dependent effective masses 
(for PV coupling) of $\p^0$(solid curve), $\p^+$(dashed curve) 
 and $\p^-$(dotted curve) at $\rho = 0.17 fm^{-3}$
 without (upper panel) and with (lower panel) vacuum correction.} 
\label{pvfig2}
\end{center}
\end{figure}
%%%%%%%%%%%%%%%%%%%%%%%%%%%%%

 It should however be mentioned that the vacuum correction part for PV 
 coupling is rather small. For loops involving heavy baryons it could
 be quite high. For detailed discussion we refer the readers to 
 \cite{Furnstahl89,Furnstahl95}. In present case we take only the
 nucleon loop in presence of the scalar mean field.\\

%%%%%%%%%%%%%%%%%%%%%%%%%%%%%%
\begin{figure}[htb]
\begin{center}
\includegraphics[scale=0.3,angle=0]{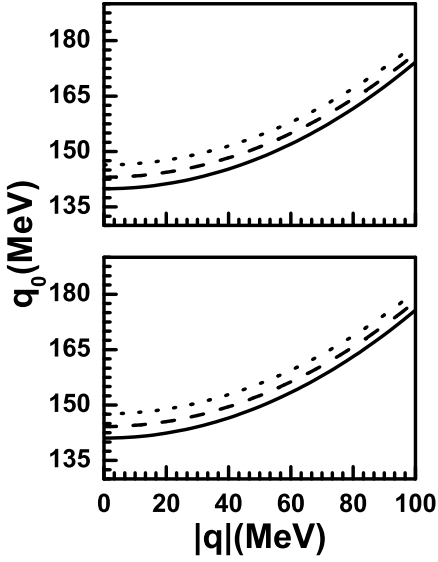}
\caption{Pion dispersion relation without (upper panel) and with (lower panel)
the effect of Dirac sea for PV coupling. The solid, dashed and dotted curves
respectively indicate the dispersion curves of $\p^0$, $\p^+$ and $\p^-$ at
$\rho = 0.17 fm^{-3}$ and $\a = 0.2$.}
\label{pvfig3}
\end{center}
\end{figure}
%%%%%%%%%%%%%%%%%%%%%%%%%%%%%

 We also present results of asymmetry parameter dependence effective 
 masses for different charge states of pion in Fig.\ref{pvfig2} at
 normal nuclear matter density. The upper and lower panel present the 
 effective pion masses without and with vacuum correction. It can be
 observed that the asymmetry parameter dependent pion mass splitting
 is insensitive to the vacuum correction. The pion dispersions in
 medium for various charge states of pion are presented in 
 Fig.\ref{pvfig3} for PV coupling.

%%%%%%%%%%%%%%%%%%%%%% MODERN TECHNIQUE %%%%%%%%%%%%%%

\section{Modern Technique}

 In the previous sections we have discussed pion propagation in ANM
 using both the PS and PV interaction within the framework of
 non-chiral model. However, the interactions as represented by
 Eq.\ref{Lag00} and Eq.\ref{TLag}, 
 fail to describe in-medium $\pi N$ dynamics as shown in \cite{Matsui82}.
 It was also observed that the chirally symmetric model (linear)
 has also various limitations \cite{Matsui82}. 
 For example, as mentioned before, it
 fails to account for the pion-nucleus dynamics in nuclear matter both in
 the PS and PV representations. In fact, it gives too strong pion nucleon
 interaction in matter which cannot be adjusted by fixing the s-wave $\pi$-$N$
 interaction in free space even in PV case. In this context the Dirac vacuum
 involving baryon loops was found to play a significant role. If one uses
 the chiral model and breaks the symmetry explicitly, the results are found
 to be very sensitive to the renormalization scheme \cite{Matsui82}. 
 In \cite{Furnstahl93} 
 it was shown that the relativistic chiral models
 with a light scalar meson appear to provide an economical marriage of 
 successful relativistic MFT and chiral symmetry. It,
 however, fails to reproduce observed properties of finite nuclei, such as 
 spin-orbit splittings, shell structure, charge densities and surface 
 energies. Since then, there has been series of attempts to construct
 a model which has the virtue of describing
 both the properties of nuclear matter and finite nuclei 
 \cite{Furnstahl87,Furnstahl93A,Furnstahl95,Furnstahl96,
 Serot97,Hu07}. Currently, the non-linear chiral effective field 
 theoretic approach seems to be
 quite successful in this respect. It might be recalled here, that, in such
 a framework, the explicit calculation of the Dirac
 vacuum is not required, rather, on the contrary, here, 
 the short distance dynamics are absorbed into
 the parameters of the theory adjusted phenomenologically by fitting empirical
 data \cite{Furnstahl89,Serot97,Hu07}. Now we proceed to calculate the
 effective pion masses in ANM in this approach.\\

 By retaining only the lowest order terms in the pion fields, one obtains
 the following Lagrangian from the chirally invariant Lagrangian\cite{Hu07} :

\bwt
\bea
\mathcal{L} &=& \bar{\Psi}(i\g_\m\partial^\m - M)\Psi + g_s\bar{\Psi}\phi_s \Psi -
g_\o \bar{\Psi}\g_\m\Phi^\m_\o \Psi - \f{g_A}{f_\p}\bar{\Psi}\g_5\g_\m\partial^\m \v{\t} \cdot \v{\Phi}_\p \Psi + \f{1}{2} \lt(\partial_\m\Phi_s\partial^\m\Phi_s - 
m^2_s\Phi^2_s\rt) \nn \\
&+& \f{1}{2}\lt(\partial_\m\v{\Phi}_\p\cdot\partial^\m\v{\Phi}_\p - m^2_\p \v{\Phi}^2\rt)
-\f{1}{4}G_{\m\n}G^{\m\n} + \f{1}{2} m^2_\o \Phi_{\o\m}\Phi^\m_\o + \mathcal{L}_{NL}
+ \dt\mathcal{L} 
%\label{}
\eea
\ewt

 The terms $\mathcal{L}_{NL}$ and $\dt\mathcal{L}$ contain,
 respectively the nonlinear terms of the meson sector and all of the 
 counterterms. The explicit expressions for $\mathcal{L}_{NL}$ and 
 $\dt\mathcal{L}$ can be found in \cite{Hu07}. \\

 It is to be noted that the meson self-energy can be found by differentiating 
 the energy density \cite{Hu07} at the two-loop level with respect to 
 the meson propagator as indicated in Fig~\ref{loop2}. One may therefore, 
 identify the $FF$, $FD$ and  $DD$ parts of the 
 self-energy with the vacuum-fluctuation(VF), Lamb-shift(LS) and 
 exchange (EX) contributions to the self-energy respectively. 
 The VF and LS terms are related to the short-range physics 
 while EX part is related to the long-range physics. 
 The detailed discussion about this short and long distance separation can be
 found in \cite{Furnstahl89,Serot97,Hu07}. 
 The diverging $FF$ part of the self-energy and LS can be expressed as a sum 
 of terms which already exists in the effective field theoretical Lagrangian 
 and can be absorbed into the counter terms. The short distance physics,
 as shown in \cite{Hu07}, while calculating exchange energies,
 are either removed by field 
 redefinitions or the coefficients are determined by fitting with the 
 empirical data. The long-range part is computed explicitly that
 produce modest corrections to the nuclear binding energy curve. This 
 can be compensated by a small adjustment of the coupling parameters. \\

%%%%%%%%%%%%%%%%%%%%% TWO LOOPS %%%%%%%%%%%%%%%%%%%%%%%%%%%%%%%
\begin{figure}[htb]
\begin{center}
\includegraphics[scale=0.40,angle=0]{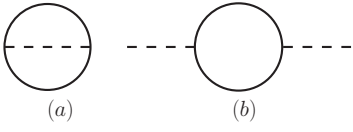}
\caption{Diagrams correspond the two loop self-energy and energy}
\label{loop2}
\end{center}
\end{figure}
%%%%%%%%%%%%%%%%%%%%%%%%%%%%%%%%%%%%%%%%%%%%%%%%%%%%%%%%%%%%%%%%

%%%%%%%%%%%%%%%%%%%%%%%%%%%%% HU %%%%%%%%%%%%%%%%%%%%%%%%%%%%%%%
\begin{figure}[htb]
\begin{center}
\includegraphics[scale=0.30,angle=0]{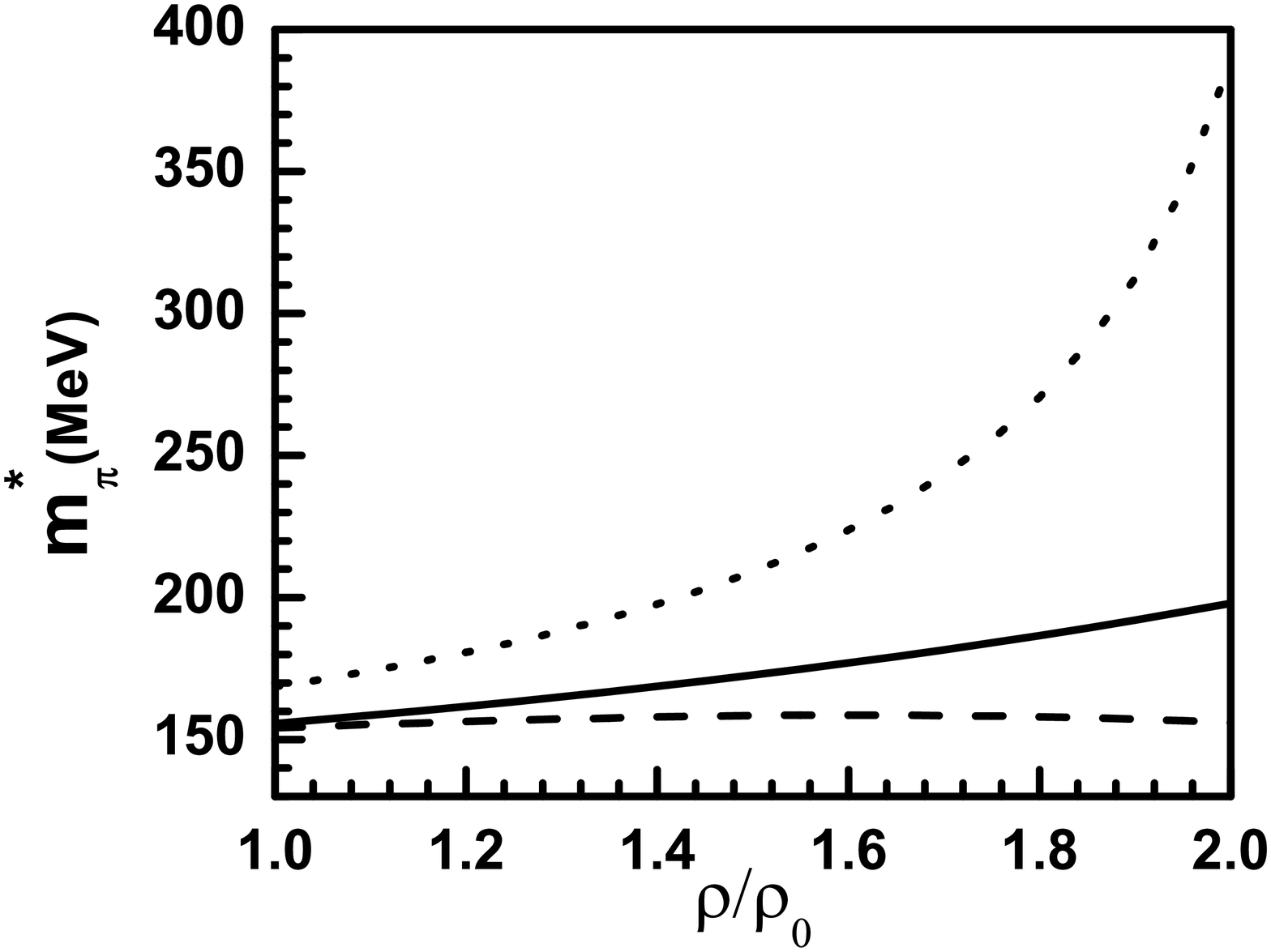}
\caption{Effective pion mass at different densities with $\alpha= 0.2 $.}
\label{hu}
\end{center}
\end{figure}
%%%%%%%%%%%%%%%%%%%%%%%%%%%%%%%%%%%%%%%%%%%%%%%%%%%%%%%%%%%%%%%%%

Recently in ref.\cite{Hu07} the exchange energy contributions of pion
has been calculated within this theoretical framework. We adopt the
same parameter set as designated by {\bf MOA} in \cite{Hu07} to calculate
the $\pi$ self-energy explicitly. The corresponding results are
presented in Fig.\ref{hu}. Here we simply depict the final results
as the expressions, at this order, for the pion self-energy and density 
dependent masses of $\pi^0$ and $\pi^\pm$ remain same as those of
Eq.\ref{pveffectivemasses} except for the
coupling parameters. Quantitatively, it is
found that, for the lower density, {\em i.e} $\rho \sim \rho_0$,
the effective masses for $\pi^-$, $\pi^0$ and $\pi^+$
states are comparable with that of PV coupling (Fig.\ref{pvfig1}), while at
higher density the mass splitting is significantly enhanced.
The charged states, {\em i.e.}   $\pi^\pm$ show stronger density
dependence compared to PV coupling.  We also observe that
the density dependence of $\pi^0$ is rather weak.    

\section{Summary and conclusion}

 In the present paper pion propagation in ANM has been
 studied within the framework of relativistic hadrodynamics 
 in presence of the scalar mean field. We start with the model developed
 in \cite{Matsui82} and present analytical results for the pion dispersion
 relations in ANM by making HNL approximation and suitable density
 expansion. Subsequently, we invoke the chirally invariant Lagrangian
 \cite{Furnstahl97,Furnstahl98} by retaining only the lowest order terms 
 in pion field and compare the results with non-chiral model calculations
 performed in section II.\\ 

 The splitting of the various charged states of pion even at normal
 nuclear matter density is found to be quite significant.
 Such mode splittings in ANM is, infact, a generic feature
 of all the isovector mesons. Therefore, it would be interesting to   
 estimate similar splitting for the $\rho$ meson and other isovector
 states. It is to be noted that the mass splitting is related
 to the pion-nucleus optical potential \cite{Weise01, Biswas06}. As
 for the dispersion relations, we restricted our calculation only to
 the time-like region which can also be extended to study the space
 like modes both for the pions and rho mesons.\\

 {\bf Acknowledgment :} The authors gratefully thank Pradip Roy  and Kausik Pal
 for valuable comments and suggestions.\\

\appendix

\section{}

After using Feynman parametrization, the term $\S^{*FF}_{PS}(q)$  in
Eq.(\ref{psff0}) can be written as

\bwt
\bea
\S^{*FF}_{PS}(q) & = & 8ig^2_\p{\m}^{2\ep}\int\f{d^Nk}{(2\p)^N}\int^1_0 dx
\lt[ \f{M^{*2}-k\cdot(k+q)}{((k+qx)^2+q^2x(1-x)-M^{*2})^2} \rt]
\nn \\
&=& \f{g^2_\p}{2\p^2}\int^1_0 dx \lt( 4\p{\m}^2\rt)^{\ep}~\f{\G(\ep)}{1-\ep} ~
\lt[ \f{M^{*2} - 3q^2x(1-x) + 2\ep q^2 x(1-x)} {\lt(M^{*2}-q^2x(1-x)\rt)^{\ep}} \rt]\nn \\
& = & \f{g^2_\p}{2\p^2}~\f{q^2}{3} + \f{g^2_\p}{2\p^2}~\f{1}{\ep}~\lt(M^{*2} -
\f{q^2}{2}\rt) -  \f{g^2_\p}{2\p^2}~\lt(M^{*2} - \f{q^2}{2}\rt) \lt(\g^\prime_E
-\ln\lt(4\p\m^2\rt)  \rt) \nn \\
& - & \f{g^2_\p}{2\p^2}~\int^1_0 dx \lt( M^{*2} - 3q^2x(1-x) \rt) \ln\lt(
M^{*2}-q^2x(1-x) \rt) \label{psff2}
\eea
\ewt

 Here $\ep = 2-\f{N}{2}$ and $\m$ is an arbitrary scaling
 parameter. $\g_E$ is the
 Euler-Mascheroni constant and $\g^\prime_E = (\g_E -1)$. The imaginary
 part of $\S^{*FF}_{PS}(q)$ can easily be found by simply replacing 
 $\ln\lt(M^{*2}-q^2x(1-x)\rt)$ with
 $\ln\lt(M^{*2}-q^2x(1-x)-i\eta\rt)$ where $\eta$ is an arbitrarily small
 parameter and the term $i\eta$ comes from the denominator of $G^F_i$
 when Feynman parametrization is performed considering $i\z$ in the 
 denominator of the propagator.\\

 Here the term $\ln\lt(M^{*2}-q^2x(1-x)\rt)$ has branch cut only for
 $M^{*2}-q^2x(1-x)<0$ and it begins at $q^2=4M^{*2}$ {\em i.e.} the threshold
 condition for nucleon-antinucleon pair production. So the limit of
 x-integration changes from $(0,1)$ to
 $(\f{1}{2}-\f{1}{2}\a,\f{1}{2}+\f{1}{2}\a)$ where
 $\a=\sqrt{1-\f{4M^{*2}}{q^2}}$ and we used ${\rm
 Im}\ln\lt(Z-i\eta\rt) = -\p$.
 Now,

\bea
\int^{\f{1}{2}+\f{1}{2}\a}_{\f{1}{2}-\f{1}{2}\a}dx~\th\lt( q^2-4M^{*2}\rt)
& = & \sqrt{1-\f{4M^{*2}}{q^2}} ~\th\lt(q^2-4M^{*2}\rt) \nn \\
& & \label{A1}
\eea

 Now the imaginary part of $\S^{*FF}_{PS}(q)$ is,

\bea
{\rm Im}~\S^{*FF}_{PS}(q) & = &
-\f{g^2_\p}{2\p^2}\int^1_0~dx~\lt(M^{*2}-3q^2x(1-x)\rt) \nn \\
& \times & {\rm Im}\lt[\ln\lt(M^{*2}-q^2x(1-x)-i\eta\rt)\rt] \nn \\
&=&-\f{g^2_\p}{4\p} \lt[q\sqrt{q^2-4M^{*2}}\rt]~\th\lt(q^2-4M^{*2}\rt) \nn \\
& & \label{psimaginary}
\eea

 It is clear from the expression of Eq.(\ref{psff2}) that the second term is
 divergent in the limit $\ep \ra 0$ (as $N \ra 4$). To remove the
 divergences we need to add the counterterms \cite{Matsui82} in the
 original Lagrangian interaction . The diverging part of Eq.(\ref{psff2}) is

\bea {\cal D}_{PS} &=& \f{g^2_\p}{2\p^2}~\f{1}{\ep}~\lt(M^{*2} - \f{q^2}{2}\rt)
\nn \\
&=&  \f{g^2_\p}{2\p^2}\lt[\f{M^2}{\ep} - \f{2}{\ep}Mg_s\phi_0 +
\f{1}{\ep}g^2_s\phi^2_0 - \f{q^2}{2\ep}\rt] \label{psdiv0} \eea

 In Eq.(\ref{psdiv0}) we substitute  the effective nucleon mass
 $M^*=(M-g_s\phi_0)$ where $M$ is the nucleon mass and $\phi_0$ is the vacuum
 expectation value of the scalar field $\phi_s$. The expression given in
 Eq.(\ref{psdiv0}) tells us that we need to be added four counter terms
 \cite{Matsui82} with the original interaction Lagrangian to remove the
 divergences from $\S^{*FF}_{PS}$. Therefore the counter term Lagrangian
 \cite{Matsui82} is denoted as

\bwt
\bea
{\cal L}_{CT}= -~\f{1}{2!}{\b}_1{\Phi}_\p\cdot\lt(\partial^2+m^2_\p
\rt)\cdot {\Phi}_\p + \f{1}{2!}{\b}_2{\Phi}^2 +
\f{1}{2!}{\b}_3{\phi}_s{\Phi}^2_\p +\f{1}{2!2!}{\b}_4{\phi}^2{\Phi}^2_\p
\label{psct0}
\eea
\ewt

 The value of the counterterms $\b_1$, $\b_2$, $\b_3$ and  $\b_4$ are 
 determined by imposing the appropriate renormalization conditions.

\bea \b_1 &=& \lt(\f{\partial\S^{FF}_{PS}(q)}{\partial q^2} \rt)_{q^2=m^2_\p}
\label{psct1}\\
\b_2 &=& \lt( \S^{FF}_{PS} \rt)_{q^2=m^2_\p}
\label{psct2}\\
\b_3 &=& -g_s\lt( \f{\partial\S^{FF}_{PS}(q)}{\partial M} \rt)_{q^2=m^2_\p}
\label{psct3}\\
\b_4 &=& -\dt\l + g^2_s\lt( \f{\partial^2\S^{FF}_{PS}(q)}{\partial M^2}
\rt)_{q^2=m^2_\p} \label{psct4} \eea

 Here $\b_1$ and $\b_2$ are the wave function and pion mass renormalization
 counterterms respectively while $\b_3$ and $\b_4$ are the vertex 
 renormalizaton counterterms  for the $\phi_s\Phi^2_\p$ vertex and 
 $\phi^2_s\Phi^2_\p$ vertex respectively. The conditions of 
 Eq.(\ref{psct1})-(\ref{psct2}) implies that the
 pion propagator $G_\p = [q^2 - m^2_\p - \S^{*R}_{PS}(q)]^{-1}$ reproduces the
 physical mass of pions in free space. The counterterm $\b_4$ determines the
 strength of coupling of the $\phi^2_s\Phi^2_\p$ vertex. In fact
 $\S^{FF}_{PS}(q)$ is found by simply replacing $M^*$ with $M$ in
 Eq.(\ref{psff2}). We can set $\dt\l=0$ to minimize the effects of many-body
 forces in the nuclear medium \cite{Matsui82} which is consistent with the
 renormalization scheme for scalar meson. Using the conditions given in
 Eqs.(\ref{psct1})-(\ref{psct4}) the following results are found :

\bea
\b_1 &=& \f{g^2_\p}{2\p^2} \lt[ \f{1}{3} \right. \left. -\f{1}{2}
\lt(\f{1}{\ep}-\g^\prime_E+\ln(4\p\m^2) \rt) \rt]\nn \\
&+& \f{g^2_\p}{2\p^2} \lt[ \int^1_0 dx~3x(1-x)\ln\lt( M^2-m^2_\p x(1-x)\rt)\rt]\nn \\
&+&  \f{g^2_\p}{2\p^2} \lt[ \int^1_0 dx \right. \left. \f{M^2x(1-x)-3m^2_\p
x^2(1-x)^2}{M^2-m^2_\p x(1-x)}  \rt] \nn \\
& & \label{psbeta1} \\
%%-------------------------------%%----------------------------
& & \nn \\
%%-------------------------------%%----------------------------
\b_2 &=& \f{g^2_\p}{2\p^2}\lt[ \f{m^2_\p}{2}+\lt(M^2-\f{m^2_\p}{3}\rt)\right.
\left.\lt(\f{1}{\ep}-\g^\prime_E+\ln(4\p\m^2) \rt) \rt]\nn \\
&-&  \f{g^2_\p}{2\p^2} \lt[ \int^1_0 dx~\lt(M^2-3m^2_\p x(1-x) \rt) \rt. \nn \\
&\times & \lt. \ln(M^2-m^2_\p x(1-x))\rt] \label{psbeta2} \\
%%--------------------------------%%-----------------------------
& & \nn  \\
%%-------------------------------%%------------------------------
\b_3 &=& \f{g^2_\p}{2\p^2} \lt[-g_s(2M)\lt( \f{1}{\ep}-\g^\prime_E+\ln(4\p\m^2)\rt) \rt]\nn \\
&+& \f{g^2_\p}{2\p^2}\lt[ g_s(2M)\int^1_0 dx \ln(M^2-m^2_\p x(1-x)) \rt]\nn \\
&+&\f{g^2_\p}{2\p^2}\lt[g_s(2M)\int^1_0 dx \right. \left. \lt(\f{M^2-3m^2_\p
x(1-x)}{M^2-m^2_\p x(1-x)}\rt) \rt] \nn \\
& & \label{psbeta3} \\
%%-------------------------------%%---------------------------------
& & \nn \\
%%------------------------------%%----------------------------------
\b_4 &=& -\f{g^2_\p}{2\p^2}6g^2_s + \f{g^2_\p}{2\p^2} \lt[2 g_s\lt(
\f{1}{\ep}-\g^\prime_E+\ln(4\p\m^2) \rt) \rt]\nn \\
&-& \f{g^2_\p}{2\p^2}\lt[2g^2_s\int^1_0 dx \ln(M^2-m^2_\p x(1-x)) \rt]\nn \\
&-& \f{g^2_\p}{2\p^2}\lt[2g^2_s\int^1_0 dx \right. \left. \f{4M^2m^2_\p
x(1-x)}{\lt(M^2-m^2_\p x(1-x) \rt)^2}\rt]\nn \\
& & \label{psbeta4}
\eea

 Now the renormalized $\S^{*FF}_{PS}(q)$ is

\bea
\S^{*R}_{PS}(q,m_\p) &=& \S^{*FF}_{PS}(q) - \b_1(q^2-m^2_\p) \nn \\
& - & \b_2-\b_3\phi_0 -\f{1}{2}\b_4{\phi}^2_0 \label{psrenorm0}
\eea

 Substituting $\S^{*FF}_{PS}(q)$ from Eq.(\ref{psff2}) and $\b_1$, $\b_2$,
 $\b_3$, $\b_4$ from Eqs.(\ref{psbeta1})-(\ref{psbeta4}) in 
 Eq.(\ref{psrenorm0}) it is found that divergences in
 $\S^{*FF}_{PS}(q)$ are completely eliminated by the counterterms. 
 After simplification $\S^{*R}_{PS}(q,m_\p)$ reduces to

\bwt
\bea
\S^{*R}_{PS}(q,m_\p) &=& \f{g^2_\p}{2\p^2}\lt[-3(M^2-M^{*2})+\right.
\left.(q^2-m^2_\p)\lt(\f{1}{6}+\f{M^2}{m^2_\p}\rt) \right. \left. -
2M^{*2}\ln\lt(\f{M^*}{M}\rt)+\f{8M^2(M-M^*)^2}{(4M^2-m^2_\p)}\right.
\nn \\
&-& \left. \f{2M^{*2}\sqrt{4M^{*2}-q^2}}{q} \right. \left.
\tan^{-1}\lt(\f{q}{\sqrt{4M^{*2}-q^2}} \rt) \right. +\left.
\f{2M^2\sqrt{4M^2-m^2_\p}}{m_\p} \right. \left.
\tan^{-1}\lt(\f{m_\p}{\sqrt{4M^2-m^2_\p}} \rt) \right.
\nn \\
&+& \left. \lt( (M^2-M^{*2})+\f{m^2_\p(M-M^*)^2}{(4M^2-m^2_\p)} \right. \left.
+ \f{M^2}{m^2_\p}(q^2-m^2_\p) \rt)  \right. \left.
\f{8M^2}{m_\p\sqrt{4M^2-m^2_\p}} \right. \left.
\tan^{-1}\lt(\f{m_\p}{\sqrt{4M^2-m^2_\p} }\rt) \right.
\nn \\
&+& \left. \int^1_0 dx~3x(1-x)q^2 \right. \left. \ln\lt(
\f{M^{*2}-q^2x(1-x)}{M^2-m^2_\p x(1-x)} \rt)\right] \label{psrenorm1}
\eea
\ewt

\section{}

 After Feynman parametrization Eq.(\ref{pvff0}) reduces to

\bwt
\bea
\S^{*FF}_{PV}(q) &=& 8i\lt(\f{f_\p}{m_\p}\rt)^2 \m^{2\ep}\int\f{d^Nk}{(2\p)^N}\int^1_0 dx \lt[\f{\lt(M^{*2}+q^2x(1-x)+k^2\rt)q^2-2(k\cdot q)^2} {((k+qx)^2+q^2x(1-x)-M^{*2})^2}\rt] \nn \\
&=& -\f{q^2}{2\p^2}\lt(\f{f_\p}{m_\p}\rt)^2\int^1_0dx \lt(4\p\m^2 \rt)^\ep
\G(\ep)\lt[\f{2M^{*2}}{\lt(M^{*2}-q^2x(1-x)\rt)^\ep} \rt] \nn \\
%%%
&=&\f{q^2}{2\p^2}\lt(\f{f_\p}{m_\p}\rt)^2\lt[2M^{*2}\lt(\g_E-\ln\lt(4\p\m^2\rt) \rt)\rt]
+ \f{q^2}{2\p^2}\lt(\f{f_\p}{m_\p}\rt)^2 \lt[2M^{*2}\int^1_0 dx~\ln\lt(M^{*2}-q^2x(1-x)\rt)\rt] \nn \\
&-& \f{q^2}{2\p^2}\lt(\f{f_\p}{m_\p}\rt)^2\lt[\f{2M^{*2}}{\ep}\rt]
\label{pvff1}
\eea
\ewt

 The imaginary part of $\S^{*FF}_{PV}(q)$ can be found as

\bea
{\rm Im}~\S^{*FF}_{PV}(q) &=& -\lt(\f{f_\p}{m_\p}\rt)^2 \nn \\
& \times & \lt[\f{q}{\p}2M^{*2}\sqrt{q^2-4M^{*2}} \rt]~
\th\lt( q^2-4M^{*2}\rt) \nn \\
& & \label{pvimaginary}
\eea

 It is clear from Eq.(\ref{pvimaginary}) that ${\rm Im}\S^{*FF}_{PV}(q)$
 vanishes for $q^2<4M^{*2}$. With the same argument as stated for $PS$ 
 coupling, we excluded the imaginary part. The diverging part of 
 $\S^{*FF}_{PV}(q)$ is

\bea {\cal D}_{PV} =
-~\f{q^2}{2\p^2}\lt(\f{f_\p}{m_\p}\rt)^2\lt[\f{2M^{*2}}{\ep}\rt]
\label{pvdiv0}\eea

 Here we use simple subtraction method to remove the divergence. So, the finite
 FF part of the self-energy is

\bea
\S^{*R}_{PV}(q)&=& \S^{*FF}_{PV}(q) -\S^{*FF}_{PV}(m_\p) \nn \\
&=& \f{q^2}{2\p^2}\lt(\f{f_\p}{m_\p}\rt)^2 \nn \\
& \times & \lt[2M^{*2}\int^1_0dx~\ln\lt(\f{M^{*2}-q^2x(1-x)}{M^{*2}-m^2_\p x(1-x)}\rt) \rt]\nn \\
& &\label{pvrenorm1}
\eea

\end{document}